\documentclass[floatfix,superscriptaddress,twocolumn,aps,pre]{revtex4-2}

\usepackage{graphicx}
\usepackage{bm}
\usepackage{physics}
\usepackage[mathlines]{lineno}
\usepackage{setspace}
\usepackage[utf8]{inputenc}
\usepackage[english]{babel}
\usepackage[T1]{fontenc}
\usepackage[dvipsnames]{xcolor}
\usepackage{amssymb}
\usepackage{amsmath}
\usepackage{xcolor}
\usepackage{chemist}
\usepackage{graphicx}
\graphicspath{ {./images/} }
\usepackage{chemformula}

\begin{document}

\title{Shock-driven amorphization and melt in Fe$_2$O$_3$}

\author{C\'eline Cr\'episson}
\email{celine.crepisson@physics.ox.ac.uk}
\affiliation{Department of Physics, Clarendon Laboratory, University of Oxford, Parks Road, Oxford OX1 3PU, UK}

\author{Alexis Amouretti}
\affiliation{Sorbonne Universit\'e, CNRS, Mus\'eum National d’Histoire Naturelle, Institut de Min\'eralogie, de Physique des mat\'eriaux et de Cosmochimie, UMR7590, Paris, France}
\affiliation{Graduate School of Engineering, Osaka University, Suita, Osaka 565-0871, Japan}

\author{Marion Harmand}
\affiliation{Sorbonne Universit\'e, CNRS, Mus\'eum National d’Histoire Naturelle, Institut de Min\'eralogie, de Physique des mat\'eriaux et de Cosmochimie, UMR7590, Paris, France}
\affiliation{PIMM, Arts et Metiers Institute of Technology, CNRS, Cnam,
HESAM University, 151 boulevard de l’H\^opital, 75013 Paris, France}

\author{Chryst\`ele Sanloup}
\affiliation{Sorbonne Universit\'e, CNRS, Mus\'eum National d’Histoire Naturelle, Institut de Min\'eralogie, de Physique des mat\'eriaux et de Cosmochimie, UMR7590, Paris, France}

\author{Patrick Heighway}
\affiliation{Department of Physics, Clarendon Laboratory, University of Oxford, Parks Road, Oxford OX1 3PU, UK}

\author{Sam Azadi}
\affiliation{Department of Physics, Clarendon Laboratory, University of Oxford, Parks Road, Oxford OX1 3PU, UK}

\author{David McGonegle}
\affiliation{AWE, Aldermaston, Reading, RG7 4PR, United Kingdom}

\author{Thomas Campbell}
\affiliation{Department of Physics, Clarendon Laboratory, University of Oxford, Parks Road, Oxford OX1 3PU, UK}

\author{David Alexander Chin}
\affiliation{University of Rochester Laboratory for Laser Energetics, Rochester, NY, USA}

\author{Ethan Smith}
\affiliation{University of Rochester Laboratory for Laser Energetics, Rochester, NY, USA}

\author{Linda Hansen}
\affiliation{University of Rochester Laboratory for Laser Energetics, Rochester, NY, USA}

\author{Alessandro Forte}
\affiliation{Department of Physics, Clarendon Laboratory, University of Oxford, Parks Road, Oxford OX1 3PU, UK}

\author{Thomas Gawne}
\affiliation{Department of Physics, Clarendon Laboratory, University of Oxford, Parks Road, Oxford OX1 3PU, UK}

\author{Hae Ja Lee}
\affiliation{SLAC National Accelerator Laboratory, 2575 Sand Hill Rd, Menlo Park, CA 94025, USA}

\author{Bob Nagler}
\affiliation{SLAC National Accelerator Laboratory, 2575 Sand Hill Rd, Menlo Park, CA 94025, USA}

\author{YuanFeng Shi}
\affiliation{Department of Physics, Clarendon Laboratory, University of Oxford, Parks Road, Oxford OX1 3PU, UK}

\author{Guillaume Fiquet}
\affiliation{Sorbonne Universit\'e, CNRS, Mus\'eum National d’Histoire Naturelle, Institut de Min\'eralogie, de Physique des mat\'eriaux et de Cosmochimie, UMR7590, Paris, France}

\author{François Guyot}
\affiliation{Sorbonne Universit\'e, CNRS, Mus\'eum National d’Histoire Naturelle, Institut de Min\'eralogie, de Physique des mat\'eriaux et de Cosmochimie, UMR7590, Paris, France}

\author{Mikako Makita}
\affiliation{European XFEL, GmbH, 22869 Schenefeld, Germany}

\author{Alessandra Benuzzi-Mounaix}
\affiliation{LULI, École Polytechnique, CNRS, CEA, UPMC, Palaiseau, FRANCE}

\author{Tommaso Vinci}
\affiliation{LULI, École Polytechnique, CNRS, CEA, UPMC, Palaiseau, FRANCE}

\author{Kohei Miyanishi}
\affiliation{RIKEN SPring-8 Center, Hyogo 679-5148, Japan}

\author{Norimasa Ozaki}
\affiliation{Graduate School of Engineering, Osaka University, Suita, Osaka 565-0871, Japan}
\affiliation{Institute of Laser Engineering, Osaka University, Suita, Osaka 565-0871, Japan}

\author{Tatiana Pikuz}
\affiliation{Graduate School of Engineering, Osaka University, Suita, Osaka 565-0871, Japan}

\author{Hirotaka Nakamura}
\affiliation{Graduate School of Engineering, Osaka University, Suita, Osaka 565-0871, Japan}

\author{Keiichi Sueda}
\affiliation{RIKEN SPring-8 Center, Hyogo 679-5148, Japan}

\author{Toshinori Yabuuchi}
\affiliation{RIKEN SPring-8 Center, Hyogo 679-5148, Japan}
\affiliation{Japan Synchrotron Radiation Research Institute, Hyogo 679-5198, Japan}

\author{Makina Yabashi}
\affiliation{RIKEN SPring-8 Center, Hyogo 679-5148, Japan}
\affiliation{Japan Synchrotron Radiation Research Institute, Hyogo 679-5198, Japan}

\author{Justin S. Wark}
\affiliation{Department of Physics, Clarendon Laboratory, University of Oxford, Parks Road, Oxford OX1 3PU, UK}

\author{Danae N. Polsin}
\affiliation{University of Rochester Laboratory for Laser Energetics, Rochester, NY, USA}

\author{Sam M. Vinko}
\affiliation{Department of Physics, Clarendon Laboratory, University of Oxford, Parks Road, Oxford OX1 3PU, UK}
\affiliation{Central Laser Facility, STFC Rutherford Appleton Laboratory, Didcot OX11 0QX, UK}

\date{\today}

\begin{abstract}
We present measurements on Fe$_2$O$_3$ amorphization and melt under laser-driven shock compression up to 209(10)~GPa via time-resolved \emph{in situ} x-ray diffraction. At 122(3)~GPa, a diffuse signal is observed indicating the presence of a non-crystalline phase. Structure factors have been extracted up to 182(6)~GPa showing the presence of two well-defined peaks. A rapid change in the intensity ratio of the two peaks is identified between 145(10) and 151(10)~GPa, indicative of a phase change. Present DFT+$U$ calculations of temperatures along Fe$_2$O$_3$ Hugoniot are in agreement with SESAME 7440 and indicate relatively low temperatures, below 2000~K, up to 150~GPa. The non-crystalline diffuse scattering is thus consistent with the - as yet unreported - shock amorphization of  Fe$_2$O$_3$ between 122(3) and 145(10)~GPa, followed by an amorphous-to-liquid transition above 151(10)~GPa. Upon release, a non-crystalline phase is observed alongside crystalline $\alpha$-Fe$_2$O$_3$. The extracted structure factor and pair distribution function of this release phase resemble those reported for Fe$_2$O$_3$ melt at ambient pressure.
\end{abstract}

\maketitle

\section{Introduction}
Crystalline and molten iron oxides are of particular interest for planetary and material sciences, given that the iron-rich outer core of the Earth contains up to 5$\%$ O, in addition to other light elements (Ni, S, Si, C, H)~\cite{hirose21}. The study of iron oxide melts is thus important for the understanding of the Fe-O bonding environment within the outer core. Changes in Fe-O bonding have been proposed to be at the origin of a possible layering of the Earth outer core, of relevance to our understanding and modelling the Geodynamo~\cite{ozawa}. The behaviour of iron oxides under pressure has so far proven to be extremely rich with the observation of new stoichiometry under pressure~\cite{lavina,lavina2015,Sinmyo} and both electronic and structural transitions. In particular, under static compression, Fe$_2$O$_3$ experiences a high-spin to low-spin, and a Mott insulator-to-metal, transition at around 50-60~GPa~\cite{pasternak,greenberg,badro,sanson}. Five Fe$_2$O$_3$ phases have been reported with increasing pressure~\cite{bykova}, all present along Fe$_2$O$_3$ Hugoniot based on temperature estimates from SESAME 7440 equation of state. However, a recent study showed that Fe$_2$O$_3$ behaves differently under laser-driven shock compression: only one isostructural phase transition from $\alpha$-Fe$_2$O$_3$ to $\alpha^\prime$-Fe$_2$O$_3$ is observed at $\sim$ 50-62~GPa, associated with a low-spin to high-spin transition, and possibly a further Mott transition~\cite{Alexis}. The behaviour of Fe$_2$O$_3$ above 116~GPa has, however, not yet been investigated under either dynamic or static compression, and its melting curve as well as its melting point along the Hugoniot remain unknown. 

In this paper we report \emph{in situ} x-ray diffraction measurements of laser-driven shock compressed Fe$_2$O$_3$ between 122(3)~GPa and 209(10)~GPa. Results indicate the appearance of a non-crystalline phase, and structure factor and pair distribution function for data under shock and upon release are calculated. Temperatures along the Hugoniot are evaluated by present DFT+$U$ calculations and SESAME equation of state 7440 for Fe$_2$O$_3$. Results are interpreted as an amorphization of Fe$_2$O$_3$ up to 145(10)~GPa prior to melting above 151(10)~GPa. Pressure-induced structural changes at the atomistic level for amorphous and molten phases are discussed.

\section{Methods}
\subsection{Experimental Methods}
As shown in Fig.~\ref{fig:figure1}, our primary diagnostic is \emph{in situ} x-ray diffraction (XRD) of laser-driven shock compressed samples, at the Matter in Extreme Conditions (MEC) endstation at the Linac Coherent Light Source (LCLS)~\cite{mec} and at the BL3 endstation at the Spring8 Angstrom Compact free electron LAser (SACLA)~\cite{sacla} X-ray Free Electron Lasers (XFEL) operating in Self-Amplified Spontaneous Emission mode. 
 
For data in transmission, acquired at LCLS, two 527~nm laser beams arriving at 20$^\circ$ to the sample normal were focused on the target down to a 300~\textmu{}m diameter focal spot. Flat-top pulses had a duration of 5 to 15~ns, with a maximum energy on the sample of 60 J. X-rays at 7.08~keV (1.751~\textup{~\AA}) probed the target at an angle of 35$^\circ$ from the normal surface and a projected x-ray spot size of 60~\textmu{}m diameter in the center of the laser drive spot. 2D XRD images were recorded on 4 quadruple ePix10k detectors covered with a 50~\textmu{}m thick Al filter and 125~\textmu{}m thick plastic filter. Azimuthal integration of the 2D images includes polarization, solid angle and filter correction as well as self-attenuation from the target. For four data points (at 167, 174, 177 and 185 GPa), only one detector was used due to technical reasons and azimuthal integration of the 2D images for those data points was performed using Dioptas software~\cite{presher} including polarization and solid angle correction. 

For data in reflection, acquired at SACLA, one 532~nm laser beam arriving at 72$^\circ$ from the sample plane was focused on the target down to a 260~\textmu{}m diameter focal spot. Flat-top pulses had a duration of 5~ns and maximum energy on the sample was 13~J. X-rays at 9 or 9.98~keV (1.378 or 1.242~\textup{~\AA}) probed the target at an angle of 18$^\circ$ from the surface plane and a a projected x-ray spot size of 40~\textmu{}m diameter overlapping the laser drive spot. 2D XRD images were recorded on a single detector with no additional filter. Azimuthal integration of the 2D images was performed using Dioptas software~\cite{presher} including polarization and solid angle correction. 

In reflection geometry the x-rays probe 2-2.5~\textmu{}m of Fe$_2$O$_3$ on the rear side (i.e. the side opposite to the laser interaction), assuming that the majority of the signal is coming from one optical path length, whereas the entire sample (8~\textmu{}m thick) is probed in transmission geometry, as shown in Fig.~\ref{fig:figure1}. Data in reflection are thus effectively more resolved in time as only the last portion of the sample traversed by the shock is probed. Additionally, data in reflection are less affected by possible ablation-related preheat~\cite{falk}.

Two kinds of targets with different layer thicknesses were used and each layers were systematically measured: CH-targets, composed of a parylene-N ablator (54 and 62~\textmu{}m thick) and a Fe$_2$O$_3$ layer (7, 8 and 10~\textmu{}m thick); and Sa-targets, composed of a parylene-N ablator (39 and 54~\textmu{}m thick), a Fe$_2$O$_3$ layer (8~\textmu{}m thick), and a sapphire window (22~\textmu{}m thick). Greater details regarding layer thicknesses for the different experiments involved can be found in the supplementary material (section I). For all targets, polycrystalline column-textured Fe$_2$O$_3$ was created by physical vapor deposition and shown to correspond to $\alpha$-Fe$_2$O$_3$~\cite{blake} by XRD, displayed in Fig. S3 of the supplementary material. A 200 nm Al coating was added on top of parylene-N ablator to limit penetration of the light in the parylene-N layer. We emphasize that no glue was used in-between the layers which were deposited on top of each other. The CH-targets provide better XRD signal as there is no attenuation or diffraction from the sapphire window but the pressure determination is more reliable for Sa-targets.  

Velocity Interferometer System for any Reflector~\cite{barker} (VISAR) was used to determine breakout time, i.e. the moment when the shock leaves the Fe$_2$O$_3$ layer, and velocity history. For Sa-target pressure was determined by using the particle velocity measured at the Fe$_2$O$_3$-sapphire interface and using impedance matching method with SESAME 7440 tables for Fe$_2$O$_3$ and SESAME 7411 tables for sapphire. The apparent particle velocity measured in the sapphire was corrected to obtain the true particle velocity~\cite{cao} following the same procedure detailed in Ref.~\cite{Alexis}. Only breakout time could be measured for CH-targets due to the loss of reflectivity of the sample and pressure was determined either by interpolating breakout time versus pressure relationship obtained from Sa-target or with the help of calibrated hydrodynamic simulations using code MULTI~\cite{ramis} further detailed in Figs. S4 and S5 of the supplementary material.

\subsection{Determination of structure factor and pair distribution function}

The structure factor was calculated using the following equation:
\begin{equation}
    S(Q) =  A\frac{I(Q)}{f(Q)^2},
\end{equation}
with $I(Q)$ the intensity of the x-ray diffraction profile, $f(Q)$ the effective electronic form factor calculated using the atomic form factor~\cite{hadju} and $A$ a normalization constant. We used the Ashcroft-Langreth formalism and calculated the Ashcroft-Langreth total structure factor as previously described~\cite{liquiddiffract}. The pair distribution function, $g(r)$, was then calculated for CH-targets with maximal $Q$ range between 5.5 and 5.9~\AA$^{-1}$. Sa-targets were discarded as their reliable Q-range is limited to $\sim4$~\AA$^{-1}$ due to the sapphire signal. We calculated the Fourier transform of the structure factor to find:  
\begin{equation}
    F(r) =  \frac{2}{\pi}\int_{0}^{Q_{max}} \sin(Qr) [S(Q)-S_\infty] \,dQ,
\end{equation}
with $r$ the radial interatomic distance and $Q_{max}$ the highest available scattering momentum.
The pair distribution function is then given by
\begin{equation}
g(r) =  \frac{n(r)}{n_0}\ = 1 + \frac{F(r)}{ 4\pi n_0},
\end{equation}
with $\ n(r)$  the atomic density at a distance $r$ from a given atom and $n_0$ the average atomic density at experimental conditions. In our case we determine the density at a given pressure using SESAME 7440 tables. To determine $g(r)$  we use an iterative process~\cite{eggert,sanloup}. We minimize the background in $g(r)$ for the smallest interatomic distances ($\ r < r_{min}$) where no real signal is expected:
\begin{align*}
&r <r_{min} \Rightarrow g(r) = 0 \Rightarrow F(r) = -4\pi r n_0 \\
&\Delta F(r) = F(r) + 4\pi r n_0,
\end{align*}
where $\Delta F(r)$ is the error on $F(r)$. $\Delta F(r)$ is then evaluated and removed from $F(r)$ to obtain the corresponding $g(r)$ from which a new $S(Q)$ can be calculated, and the cycle repeated. 

\subsection{Computational Methods}
Temperatures along the Fe$_2$O$_3$ Hugoniot were calculated using density functional theory with a Hubbard $U$ parameter (DFT+$U$) \cite{Hohenberg,Kohn,Anisimov,Liechtenstein,Dudarev,Cococcioni}. Calculations were performed using the v7.2 Quantum-Espresso suite of codes \cite{QE,QE2}. Scalar-Relativistic ultrasoft pseudopotential (PP) \cite{UPF} generated by PBEsol exchange-correlation functionals \cite{PBEsol} were used for Fe and O. The effective Hubbard $U$ parameter was used for the Fe-$3d$ orbitals, with the initial occupations given by the PP. We used a kinetic energy cutoff of the plane-wave basis set of 100~Ry and an augmentation charge energy cutoff of 800~Ry. The calculations were carried out using a 12$\times$12$\times$12 $\mathbf{k}$-point grid.  Our DFT$+U$ calculations using $U=5$~eV predict that at the ambient condition, the energy band gap and the magnetic moment per iron atom are 2.075~eV and 4.41~$\mu_B$, respectively. The corresponding experimental values are 2.14~eV~\cite{Benjelloun} and 4.6~$\mu_B$~\cite{Coey}.

The equation of state (EOS) is obtained by performing DFT calculations at different densities in which the temperature effects were included using the quasi-harmonic approximation~\cite{Baroni}. The lattice dynamics and phonon spectra used in the quasi-harmonic approximation were determined by density functional perturbation theory~\cite{Baroni}. The phonon spectra were obtained using a $2\times2\times2$ $\mathbf{q}$-point mesh.
The Hugoniot pressure and temperature, $P_H$ and $T_H$, are given by the Rankine-Hugoniot (RH) equation $\frac{1}{2}P_H[V_0 - V_H] = E_H - E_0$, where $V_0$ and $E_0$ are the volume and energy of the system at ambient conditions. The (RH) equation is solved using the EOS data by varying the temperature at a fixed volume until the RH condition is satisfied.

\section{Results}

\begin{figure*}
\includegraphics[width=\textwidth]{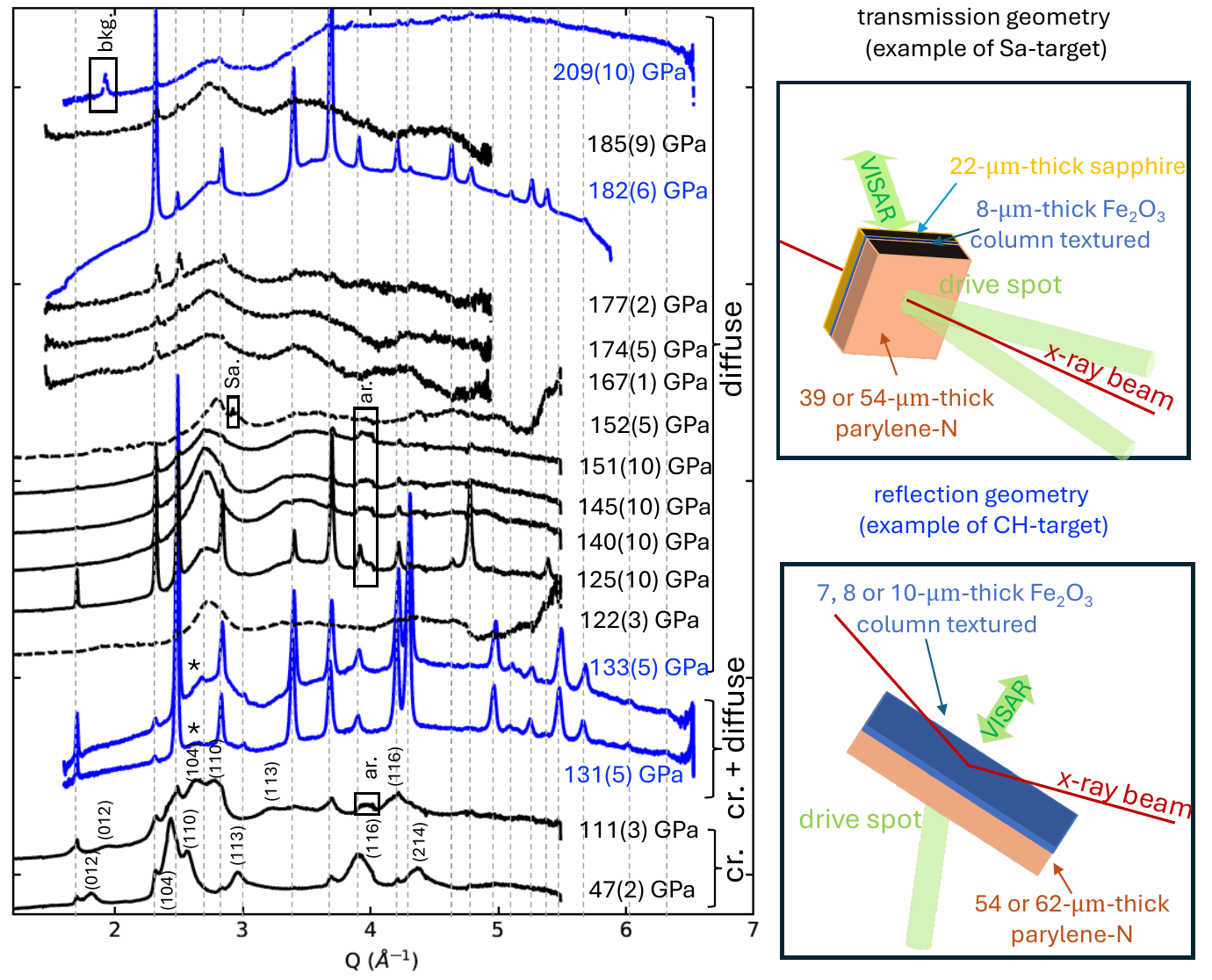}
\caption{{\bf X-ray diffraction measurements of Fe$_2$O$_3$ under shock compression.} Azimuthally integrated XRD profiles under shock showing crystalline (cr.) Fe$_2$O$_3$ identified as $\alpha$-Fe$_2$O$_3$ at 47(2) GPa and $\alpha^\prime$-Fe$_2$O$_3$ at 111(3) GPa with peak indexation reproduced from previous study~\cite{Alexis}; appearance of diffuse feature at 131(5) and 133(5) GPa with remaining crystalline peaks of $\alpha^\prime$-Fe$_2$O$_3$ (indicated by *); and fully non-crystalline Fe$_2$O$_3$ from 122(3) to 209(10) GPa. Examples of 2D image plates are shown in Figs. S6 and S7. Solid lines correspond to CH-target and dashed lines to Sa-targets. Measurements acquired in transmission geometry are shown in black, and in reflection geometry in blue. Vertical dashed lines indicate the positions of $\alpha$-Fe$_2$O$_3$ ambient peaks~\cite{blake} present before breakout time.
Spurious features in the data are indicated by: (Sa.) for a sapphire peak, (ar.)  XRD integration artifacts, and (bkg.) for a background peak which was also observed on the same sample prior to the shock (Fig. S3). Only one detector could be used for data at 167, 174, 177 and 185 GPa due to technical reasons. Patterns are scaled for clarity.}
\label{fig:figure1}
\end{figure*}

We first present the complete Fe$_2$O$_3$ diffraction dataset obtained in transmission and reflection geometries, and examine the series of pressure-induced changes to the structure factor between 47(2) and 209(10)~GPa. A waterfall plot of the azimuthally integrated diffraction patterns is provided in Fig.~\ref{fig:figure1}. As has already been shown~\cite{Alexis}, Fe$_2$O$_3$ remains crystalline under shock compression up to 116~GPa. Above 131(5)~GPa, we observed the onset of diffuse diffraction features for both the transmission and reflection geometries. The diffuse features at 2.7-2.8~\AA$^{-1}$ are first observed at 122(3) GPa in transmission and at 131(5) GPa in reflection. The second diffuse feature at 3.4-3.6~\AA$^{-1}$ becomes visible at 125(10)~GPa in transmission geometry. These diffuse features persist up to the maximum pressure achieved in our experiments of 209(10)~GPa .

In reflection geometry at 131(5) and 133(5)~GPa, the diffuse feature is observed in coexistence with crystalline high pressure $\alpha^\prime$-Fe$_2$O$_3$~\cite{Alexis} in contrast to transmission geometry, where no crystalline phase is visible from 122(3)~GPa onwards. In reflection geometry only 2-2.5~$\mu$m on the rear side of the sample is probed, i.e. the zone just behind the shock front. Detailed study of melting timescales in shocked Ge has shown that while the characteristic melting time at around 10 GPa above the melting pressure is sub-nanosecond, this can reach up to 7~ns at the melting point along the Hugoniot~\cite{renganathan}. It is thus possible to explain the remaining crystalline peak observed only for data acquired in reflection geometry by a delayed melting or amorphization behind the shock front, the effect of which will not be present in data acquired in transmission geometry. Differences between sensitivity of XRD in transmission and transverse geometry have also been previously discussed regarding onset of melting~\cite{McBride}.
The difference of around 10~GPa observed in the pressure at which the diffuse features appear and at which the crystalline phase disappears between measurements performed in transmission and reflection geometries could also be attributed to preheat effect inducing a slightly higher temperature in transmission than in reflection geometry for a given pressure. No preheat is identified in VISAR data indicating that even if preheat cannot be entirely discarded, it remains low.


To better understand the structure of the non-crystalline phase and its evolution under pressure, the structure factor $S(Q)$ and pair distribution function $g(r)$ were determined for samples with no crystalline peak. The baseline levels caused by emission from the ablation plasma of the laser drive was found on runs that contained only solid diffraction. A scaled version of this background was then removed from all the diffraction data before smoothing (further details for baseline removal can be found in Figs. S8 to S10). Our Q-range for data under shock compression ($\sim$5.5~\AA$^{-1}$) leads to a limited resolution for $g(r)$ with a remaining background contribution below 1.5~\AA. Therefore $g(r)$ cannot be reliably used to retrieve structure of the non-crystalline phase and is not discussed in length in the paper. We note that the limited Q-range does not affect the determination of the structure factor $S(Q)$~\cite{eggert}.

\begin{figure}
\includegraphics[width=\columnwidth]{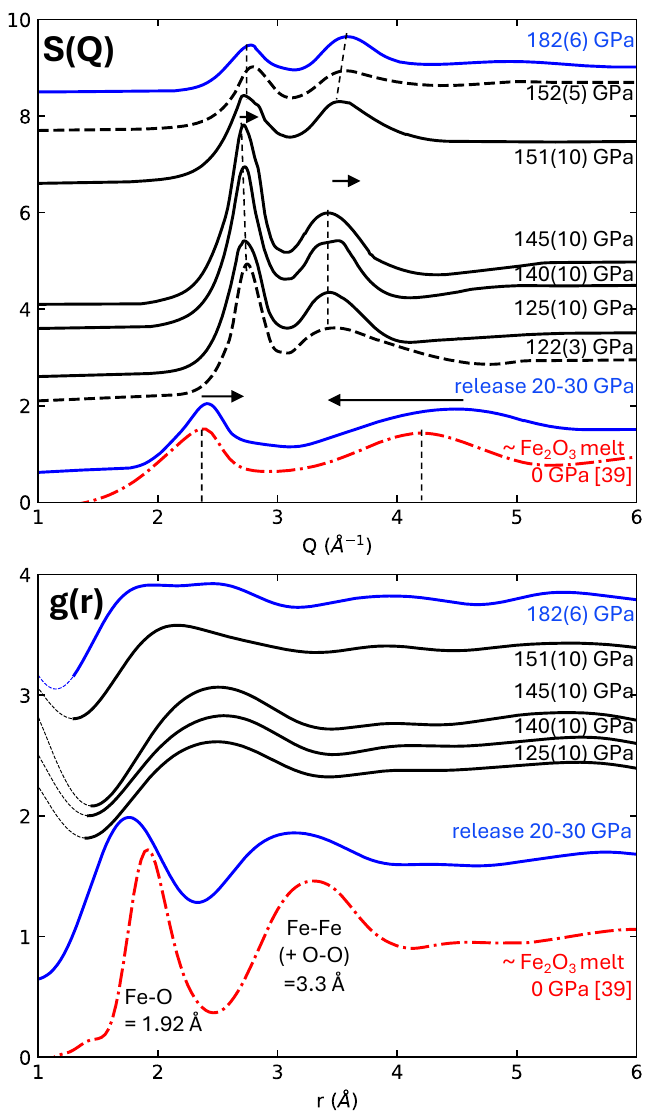}
\caption{{\bf Structure factor $S(Q)$ and $g(r)$ of Fe$_2$O$_3$.} Structure factor for non-crystalline data points under shock and upon release. Data acquired in transmission geometry are shown in black, in reflection geometry in blue and ambient in red~\cite{shi}. Dashed lines indicate Sa-targets and solid lines CH-targets. Data point upon release has a non-zero pressure as the reflection of the release wave off the parylene-N-Fe$_2$O$_3$ interface arrives on the free surface again exactly at the time when the sample is probed. Under shock, two  peaks are seen in the structure factor, the position and intensity of which are strongly different from ambient melt and from non-crystalline phase observed upon release (see vertical dotted line and arrows for guidelines). Above 145(10)~GPa a slight shift of both  peaks and an increase of the second peak intensity is observed. Despite limited resolution we can see a significant difference in $g(r)$ upon release and under shock as well as a general shift toward lower interatomic distances above 145(10) GPa.}
\label{fig:figure3}
\end{figure}

We can see in Fig.~\ref{fig:figure3} that structure factors of non-crystalline Fe$_2$O$_3$ under shock from 122(3) to 145(10)~GPa present two peaks at 2.7-2.8~\AA$^{-1}$ and 3.4-3.6~\AA$^{-1}$. The structure factors presented here differ greatly from ambient pressure Fe$_2$O$_3$ melt~\cite{shi}, as can be seen in Fig.~\ref{fig:figure3}. To our knowledge, there is no existing data on amorphous or liquid Fe$_2$O$_3$ under pressure. Currently available data on molten FeO, available up to 70 GPa and 3500 K under static compression~\cite{morard2022}, show a very different structure factor than the one measured in our experiment.
Above 145(10)~GPa significant changes are observed. In terms of the structure factor, we observe both a slight shift of peak positions up to $\sim$ +0.1~\AA$^{-1}$ and a change in the intensity ratio. The peak at 3.5~\AA$^{-1}$ becomes more prominent, and exceeds in intensity the first peak at around 2.7~\AA$^{-1}$ at the highest pressures. Moreover, both peaks significantly broaden.
These differences translate into $g(r)$ by a general shift toward lower interatomic distances above 145(10)~GPa. Between 151~GPa and 182~GPa the second peak of the structure factor increases and shifts toward higher Q values. The rapid change observed above 145(10)~GPa suggests the presence of a disorder-to-disorder phase transition.

\begin{figure}
\includegraphics[width=\columnwidth]{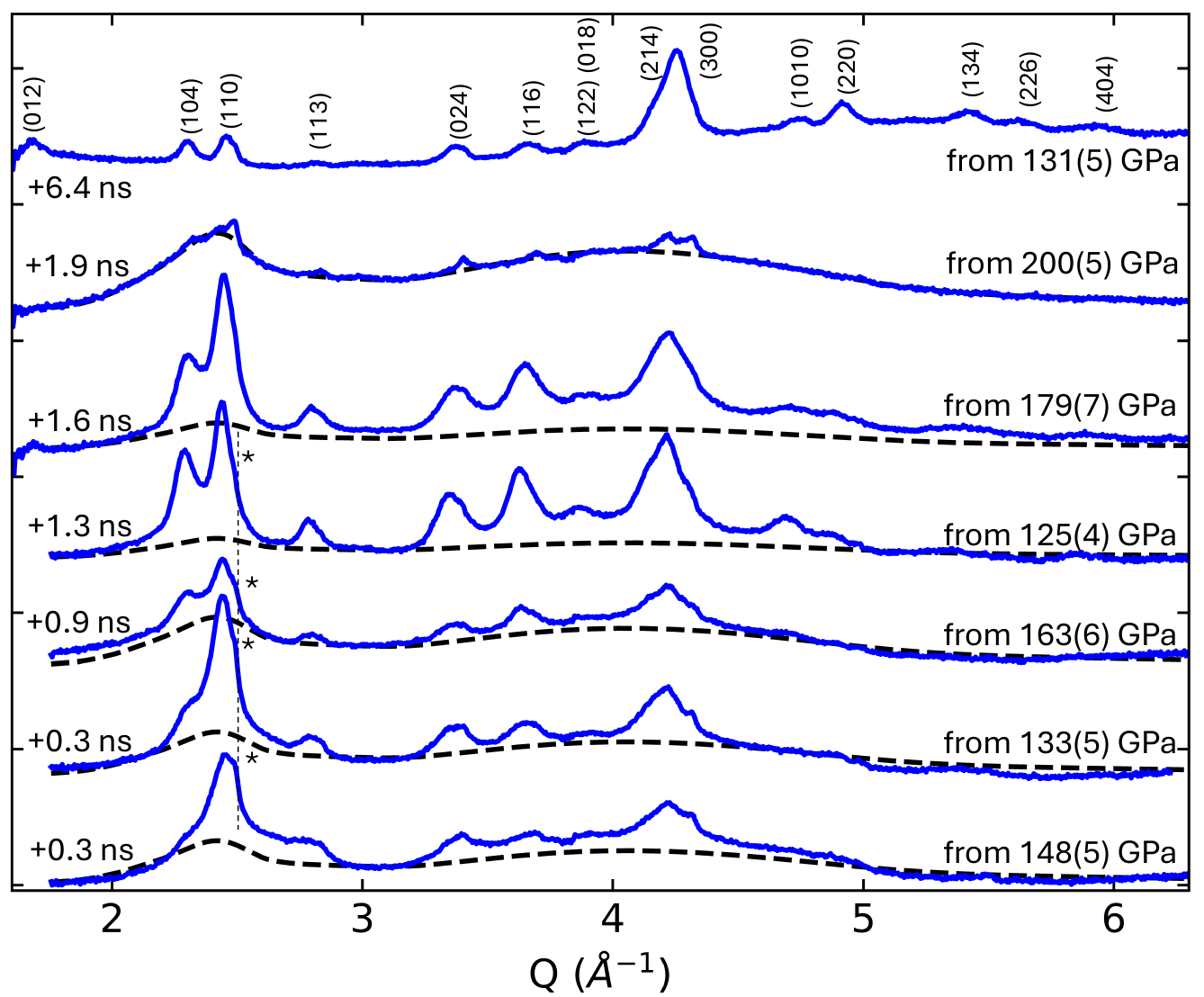}
\caption{{\bf X-ray diffraction measurements upon release.} Azimuthally integrated x-ray diffraction profiles. Data are acquired in reflection geometry on CH-target. Baseline has been subtracted to better assess the presence of a diffuse signal. The time after the shock leaves the target is indicated on the left. Crystalline peaks are seen in coexistence with diffuse features around 2.4~\AA$^{-1}$ and 4.5 ~\AA$^{-1}$ up to 0.9~ns. Very few crystalline peaks are observed for one release at 1.9~ns after breakout time after shock corresponding to 200(5)~GPa: at this timing the release wave originating at the free-surface of the Fe$_2$O$_3$ layer reflects off the parylene-N-Fe$_2$O$_3$ interface and reaches the free surface again (see Fig. S11). Diffuse signal observed at 1.9~ns after breakout time is shown as black dashed curve and adjusted by a multiplicative factor for all data up to 1.9~ns, showing the clear presence of a similar diffuse signal for release runs at 0.3 and 0.9~ns and later in time at 1.9~ns after breakout time. Peaks of $\alpha$-Fe$_2$O$_3$ are indexed~\cite{blake}. The (110) peak of $\alpha^\prime$-Fe$_2$O$_3$ indicated by *, is observed until 1.3 ns, proving that the sample is still releasing up to this timing.}
\label{fig:figure4}
\end{figure}

The XRD spectra evolution of Fe$_2$O$_3$ upon release is shown in Fig.~\ref{fig:figure4}. We consider data acquired in reflection geometry on CH-targets shocked to pressures between 125(4) and 179(7) GPa, i.e. in the pressure range that shows a non-crystalline phase in Fig.~\ref{fig:figure1}. We observe a diffuse signal at 2.4 and 4.5~\AA$^{-1}$ in coexistence with crystalline Fe$_2$O$_3$, up to approximately 0.9 ns after breakout time. Purely crystalline Fe$_2$O$_3$ is mostly observed beyond 0.9 ns. The $\alpha$-Fe$_2$O$_3$ phase is mostly present in coexistence with $\alpha^\prime$-Fe$_2$O$_3$, whose (110) peak is visible up to 1.3~ns after breakout time.
A case of particular interest is at 1.9~ns after breakout, after a shock at 200(5) GPa, the release wave originating at the free-surface of the Fe$_2$O$_3$ layer reflects off the parylene-N-Fe$_2$O$_3$ interface and reaches the free surface again. The XRD pattern taken at this time (displayed in Fig.~\ref{fig:figure4}) shows a loss of crystalline peaks and a non-crystalline signal. The diffuse signal is shown as a black dashed curve and adjusted by a multiplicative factor for all data up to 1.9~ns, showing the clear presence of a similar diffuse signal on Fig.~\ref{fig:figure4} for release runs at 0.3 and 0.9~ns and later in time at 1.9~ns after breakout time, while diffuse signal is hardly visible for release runs at 1.3 and 1.6~ns. The large proportion of non-crystalline phase for this data point may be linked to the pressure under shock, higher than the other release data points implying a different release path, although the unloading history cannot be fully understood due to the large variety of phenomena happening under release at this large timing, not included in our hydrodynamic simulations. Non-crystalline signal, similar to earlier release data points (up to 0.9~ns after breakout time) where no reflection of release wave yet occurred, can more easily be extracted on this particular data point to get $S(Q)$ and $g(r)$, displayed in Fig.~\ref{fig:figure3}. We note that extracted $S(Q)$ and $g(r)$ differ significantly from the data under shock, but are similar to data taken on molten Fe$_2$O$_3$ at ambient pressure~\cite{shi} with a slight shift linked to pressure arising from the reflection of the release wave previously described. We have thus observed a disordered phase, whose structure resembles ambient, molten Fe$_2$O$_3$~\cite{shi}, recrystallizing in time.

\section{Discussion}

\begin{figure}
\includegraphics[width=\columnwidth]{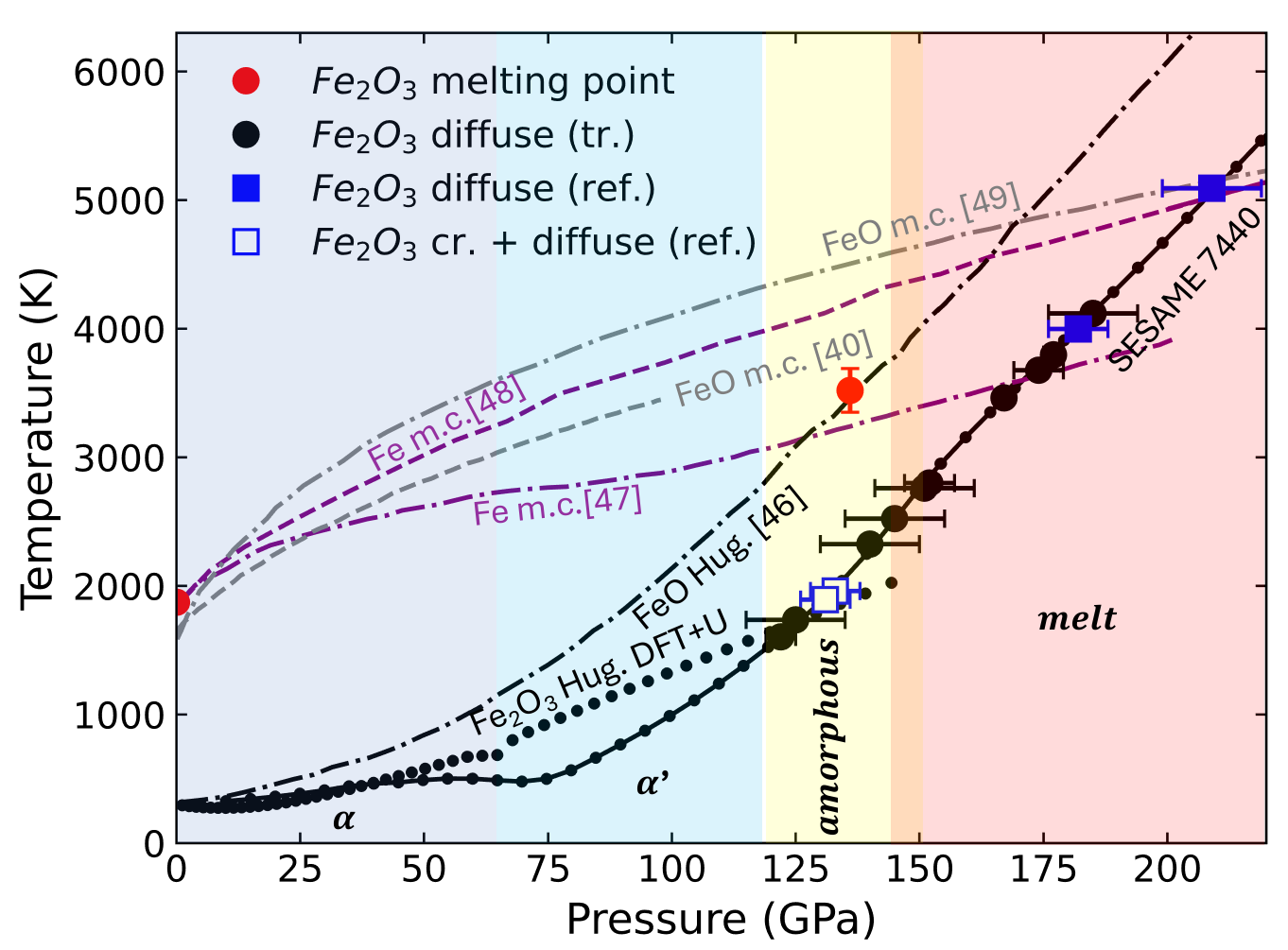}
\caption{{\bf Temperature-pressure diagram of Fe$_2$O$_3$ under shock.} 
The melting point at 0 GPa is taken from Ref.~\cite{biswas} and the assumed melting point at 136 GPa from Ref.~\cite{deng}. For comparison we also show the FeO Hugoniot (Hug.)~\cite{jeanloz}, and the melting curves (m.c.) of Fe~\cite{boehler1993,anzellini} and FeO~\cite{morard2022,dobrosavljevic}. We represent our data points with diffuse feature and in some case remaining crystals (cr.) under shock using temperature predicted by SESAME 7440. Results of DFT+$U$ calculations, based on previous shock data~\cite{Alexis}) are shown to agree with SESAME 7440. Phases observed along Fe$_2$O$_3$ Hugoniot under shock compression are summarized: $\alpha$-Fe$_2$O$_3$ stable up to approximately 50 GPa and $\alpha^\prime$-Fe$_2$O$_3$ phases, observed alone, above 80 GPa~\cite{Alexis}, amorphous phase from 122(3) to 145(10) GPa and melt from 151(10) GPa. Colours are used as a guide to distinguish phases along the Hugoniot and do not infer phases off-Hugoniot.}
\label{fig:figure2}
\end{figure}

Temperature measurements during shock compression experiments remain scarce. Moreover, shock-temperature cannot be directly extracted from Rankine-Hugoniot relations and approximations are needed to evaluate temperature along the Hugoniot~\cite{Raikes,Sharp}. For these reasons there is still a large uncertainty on temperature estimates along the Hugoniot of various materials. Temperatures along the Fe$_2$O$_3$ Hugoniot are only known from theoretical calculations or equation of state tables such as SESAME 7440~\cite{kalita}, shown in Fig.~\ref{fig:figure2}. However, because we know that under shock compression the phases are different to those observed statically~\cite{Alexis}, we have performed DFT+$U$ calculations using the phases directly observed under shock up to 116~GPa to determine the temperature along Fe$_2$O$_3$ Hugoniot. This provides an alternative method based on the most recent results to constrain the temperature range explored in our experiments. A reasonable agreement is observed between temperature predicted by SESAME 7440 and our DFT+$U$ calculations. Temperatures remain relatively low (below 1000~K up to around 80~GPa), with a change in slope at 60 GPa, linked to the isostructural phase transition reported under shock~\cite{Alexis}. 

At ambient pressure Fe$_2$O$_3$ melts at 1873~K~\cite{biswas}. Although the melting curve of Fe$_2$O$_3$ has not been measured to date, a melting temperature of 3350-3690~K at 136~GPa was estimated for $\eta$-Fe$_2$O$_3$ from first-principles molecular dynamic melting temperature calculations for FeO$_2$H$_x$~\cite{deng} assuming a linear relationship between the melting temperature and the oxygen content for a FeO$_x$ system. Fe$_2$O$_3$ could thus, possibly, melt at a lower temperature than FeO under pressure. Temperatures for data acquired at 122-133~GPa remain below the ambient pressure melting point of Fe$_2$O$_3$. We consequently attribute the diffuse features observed at this pressure to an amorphization of Fe$_2$O$_3$ under shock, although the possibility of a negative melting curve for Fe$_2$O$_3$~\cite{LULI}, which could explain the presence of a melt at 122-133~GPa along the shock Hugoniot, cannot be entirely discarded.

Solid state amorphization under shock compression has been reported on a wide range of materials~\cite{idrissi,li} including silica~\cite{fused_silica} and silicates such as plagioclase~\cite{gleason}, enstatite~\cite{hernandez} or olivine~\cite{jeanloz_1977,kim2020,shim2023}, as well as in covalently bonded solids such as Ge, Si, SiC and B$_4$C~\cite{zhao}. Amorphization has been shown to be favored by compression, shear and high strain-rates as present during shock compression. While it remains debated why and how a material becomes amorphous, amorphization is commonly linked to deformation mechanisms. It is seen either as the final step of plastic deformation or as an alternative deformation mechanism in competition with other types of deformations~\cite{idrissi,li}. 

DFT+$U$ calculations were performed to determine bulk and Young moduli of $\alpha^\prime$-Fe$_2$O$_3$ phase at 120 GPa and assess its stability with respect to the applied shear stress. The elastic constants tensor is used to obtain the bulk modulus $B$, the Young modulus $Y$, the shear modulus $G$, and the Poisson ratio $n$.
We first validated our DFT+$U$ simulations by verifying the accuracy of the derived $B$, $Y$, $G$ and $n$, for the ambient $\alpha$-Fe$_2$O$_3$ phase, which in the experiment (versus in the simulation) take values of 203(4)~GPa~\cite{liberman,schouwink} (versus 212~GPa), 220~GPa~\cite{saeki} (versus 241~GPa), 101(9)~GPa~\cite{liberman,greer} (versus 92~GPa) and 0.31~\cite{liberman} (versus 0.31), respectively. We performed the same calculations for $\alpha^\prime$-Fe$_2$O$_3$ phase at 120 GPa, and it was found that both the shear modulus and Young modulus are negative. The negative shear modulus suggests that the $\alpha^\prime$ crystal structure cannot sustain the applied shear stress, leading to a deformation or structural rearrangement. This is consistent with the observed destabilization of $\alpha^\prime$-Fe$_2$O$_3$ phase, transforming into an amorphous phase at 122(3) GPa.

We can thus conclude that above 122(3) GPa amorphous Fe$_2$O$_3$ is kinetically favorable compared to other crystalline phases, or to the further deformation of $\alpha^\prime$-Fe$_2$O$_3$ whose shear and Young moduli are found to be negative at this pressure.

The change of peak ratio intensity observed between 145(10)~GPa and 151(10)~GPa translate into $g(r)$ by a general shift toward lower interatomic distances above 145(10)~GPa, as shown in Fig.~\ref{fig:figure3}. An increase in the intensity of the second peak of the structure factor under pressure is frequent in silicate glasses and melts, and is attributed to the gradual transition from fourfold to sixfold coordinated Si~\cite{meade,gleason,morard2020,shim2023}. Nevertheless, non-crystalline Fe$_2$O$_3$ structure is unlikely to be comparable with glass and melt silicate structures under pressure, and measurements at higher Q-range will be needed to uncover the Fe-O and Fe-Fe interatomic distances and coordination numbers to fully interpret this observation. 
In our case, the peak ratio intensity rapidly changes within 6~GPa (from 145(10) to 151(10)~GPa) while both peaks broaden. This can be either interpreted as a melting of Fe$_2$O$_3$, or a rapid transition from amorphous Fe$_2$O$_3$ to another amorphous phase. The change observed here, within few GPa, seems to be in better agreement with the melting of Fe$_2$O$_3$. Indeed, changes observed so far in amorphous phases under shock are gradual, such as observed for amorphous plagioclase~\cite{gleason}. Amorphous phases are also seen to span large pressure ranges with no structural variation under shock compression, as shown for enstatite~\cite{hernandez}, but also as reported here in Fig.~\ref{fig:figure3} where the structure factors remain unchanged between 122(3) and 145(10)~GPa.
The subsequent change observed from 145(10) to 151(10)~GPa can be related to the rapid structural evolution of the melt, whose structure factor is also seen to evolve significantly as the pressure is increased further. The melting of Fe$_2$O$_3$ at 151~GPa along the Hugoniot would thus indicate a relatively low melting temperature around 2500~K. 
 
Previous laser shock data has shown that the phase transition observed along the Fe$_2$O$_3$ Hugoniot between 140 GPa and 300 GPa is accompanied by an increase in density and thus a decrease in volume~\cite{LULI}. If the observed transition is associated to melting based on the present results, this will indicate a negative Clapeyron slope for Fe$_2$O$_3$ melting curve. In which case the amorphization observed prior to melting could be explained by the Fe$_2$O$_3$ Hugoniot crossing the metastable extension of the Fe$_2$O$_3$ melting curve at around 122(3) GPa, a phenomenon already reported for quartz~\cite{hemley}. We note, however, that amorphization prior to melting under shock is also observed for enstatite~\cite{hernandez} and olivine~\cite{shim2023}, whose melting curves present a positive slope at the investigated pressures. No discontinuity is observed along the Fe$_2$O$_3$ Hugoniot between 100 and 140 GPa~\cite{Mcqueen,marsh,LULI} which could indicate that amorphous Fe$_2$O$_3$ has a density close to $\alpha^\prime$-Fe$_2$O$_3$ observed up to 116 GPa~\cite{Alexis}. 

As we show in Fig.~\ref{fig:figure3}, the structure factor and pair distribution function of the non-crystalline release phase observed at 1.9~ns after breakout is similar to that of liquid Fe$_2$O$_3$ at ambient pressure~\cite{shi}, apart from a slight shift of -0.16~\AA\ which we attribute to the pressure linked with the release wave originating at the free-surface of the Fe$_2$O$_3$ layer reflecting off the parylene-N-Fe$_2$O$_3$ interface and reaching the free surface exactly when the sample is probed. While we cannot take into account the full variety of processes occurring upon release in our hydrodynamic simulations, we are able to estimate the pressure of the system based on the observed bond lengths decrease. Assuming an isotropic compression, we estimate the pressure based on the Fe-O or Fe-Fe bond lengths upon release ($d$), the bond lengths at ambient pressure~\cite{shi} ($d_0$), and the adiabatic modulus ($K$) of liquid FeO~\cite{komabayashi2014} (as no data is available for liquid or amorphous Fe$_2$O$_3$), as given by:
\begin{equation}
P = -K \ln \left(\frac{d^3}{d_0^3}\right).
\end{equation}
We find a pressure of $\sim$30 and 20~GPa from the Fe-O and Fe-Fe bond length decreases, respectively. The diffuse features observed upon release decrease in intensity up to 0.9 ns after breakout time after shock of 125(4) to 179(7) GPa. In contrast, features from the $\alpha^\prime$-Fe$_2$O$_3$ phase persist for longer, indicating that the sample is still releasing up to around 1.3 ns, as shown in Fig.~\ref{fig:figure4}. These observations could indicate either the presence of a melt recrystallizing in time, explaining the similarity of the non-crystalline release phase with liquid from $\alpha$-Fe$_2$O$_3$ at ambient pressure~\cite{shi}, or the relaxation and the recrystallization of the amorphous phase with a structure similar to the corresponding liquid at ambient pressure. Similar results and observations were reported for forsterite upon release~\cite{kim2020}. We note that for a release after a higher pressure shock of 200(5) GPa the non-crystalline phase is dominant at 1.9~ns after breakout in one case, where the reflection of a release wave off the parylene-N-Fe$_2$O$_3$ interface reaches the Fe$_2$O$_3$ layer. This implies a need to consider the full unloading history of the sample to understand the entire behaviour of Fe$_2$O$_3$ upon release through timing.    

\section{Conclusion}
It has been recently shown that shock-compressed Fe$_2$O$_3$ does not experience any of the phase transitions reported by static compression up to 116 GPa, but instead undergoes an isostructural transition (from $\alpha$- to $\alpha^\prime$-Fe$_2$O$_3$) at 50-62~GPa~\cite{Alexis}. In the present study we show that from 122-131~GPa up to 145(10) GPa, amorphization is preferred to either deformed $\alpha^\prime$-Fe$_2$O$_3$ or to other potential phase transitions. This is similar to previous results reported for covalently (or partially-covalently)-bonded solids~\cite{zhao}. Between 145(10) and 151(10)~GPa we observed a rapid change in the peak intensity ratio in the structure factor. We attribute this to the appearance of Fe$_2$O$_3$ melt, which rapidly evolves under pressure. Further experiments at higher photon energy should make it possible to access higher Q-range and thoroughly explain these observations in terms of structural evolution of amorphous and molten phases. 

\section{Acknowledgements}
Use of the Linac Coherent Light Source (LCLS), SLAC National Accelerator Laboratory, is supported by the U.S. Department of Energy, Office of Science, Office of Basic Energy Sciences under Contract No. DE-AC02-76SF00515. Part of the experiment was performed at BL3 of SACLA with the approval of the Japan Synchrotron Radiation Research Institute (JASRI) (proposals No. 2021A8643 and No. 2023A8061), in combination with the high-power nanosecond laser of the Institute of Laser Engineering, Osaka University. A.A. and N.O. were supported by grants from Japan Society for the Promotion of Science (JSPS), KAKENHI (Grants No. 22KF0243, No. 20H00139, No. 22K18702) and Core-to-Core program (Grant No. JPJSCCA20230003). This project has received funding from the European Research Council (ERC) under the European Union’s Horizon 2020 research and innovation program (ERC PLANETDIVE grant agreement No 670787). 
This material is based upon work supported by the Department of Energy [National Nuclear Security Administration] University of Rochester "National Inertial Confinement Fusion Program" under Award Number(s) DE-NA0004144.
C.C., P.P., S.A., J.S.W and S.M.V. acknowledge support from the UK EPSRC under grants EP/P015794/1, EP/X031624/1 and EP/W010097/1. 
T.G. acknowledges support from AWE via the Oxford Centre for High Energy Density Science (OxCHEDS).
T.C. and S.M.V. acknowledge support from the Royal Society.
A.F. acknowledges support from the STFC UK Hub for the Physical Sciences on XFELs.
We thank the microscopy, x-ray diffraction and PVD platforms at IMPMC for support in producing and characterizing the $\text{Fe}_2\text{O}_3$ samples.

\bibliography{mybiblio}

\begin{thebibliography}{69}%
\makeatletter
\providecommand \@ifxundefined [1]{%
 \@ifx{#1\undefined}
}%
\providecommand \@ifnum [1]{%
 \ifnum #1\expandafter \@firstoftwo
 \else \expandafter \@secondoftwo
 \fi
}%
\providecommand \@ifx [1]{%
 \ifx #1\expandafter \@firstoftwo
 \else \expandafter \@secondoftwo
 \fi
}%
\providecommand \natexlab [1]{#1}%
\providecommand \enquote  [1]{``#1''}%
\providecommand \bibnamefont  [1]{#1}%
\providecommand \bibfnamefont [1]{#1}%
\providecommand \citenamefont [1]{#1}%
\providecommand \href@noop [0]{\@secondoftwo}%
\providecommand \href [0]{\begingroup \@sanitize@url \@href}%
\providecommand \@href[1]{\@@startlink{#1}\@@href}%
\providecommand \@@href[1]{\endgroup#1\@@endlink}%
\providecommand \@sanitize@url [0]{\catcode `\\12\catcode `\$12\catcode
  `\&12\catcode `\#12\catcode `\^12\catcode `\_12\catcode `\%12\relax}%
\providecommand \@@startlink[1]{}%
\providecommand \@@endlink[0]{}%
\providecommand \url  [0]{\begingroup\@sanitize@url \@url }%
\providecommand \@url [1]{\endgroup\@href {#1}{\urlprefix }}%
\providecommand \urlprefix  [0]{URL }%
\providecommand \Eprint [0]{\href }%
\providecommand \doibase [0]{https://doi.org/}%
\providecommand \selectlanguage [0]{\@gobble}%
\providecommand \bibinfo  [0]{\@secondoftwo}%
\providecommand \bibfield  [0]{\@secondoftwo}%
\providecommand \translation [1]{[#1]}%
\providecommand \BibitemOpen [0]{}%
\providecommand \bibitemStop [0]{}%
\providecommand \bibitemNoStop [0]{.\EOS\space}%
\providecommand \EOS [0]{\spacefactor3000\relax}%
\providecommand \BibitemShut  [1]{\csname bibitem#1\endcsname}%
\let\auto@bib@innerbib\@empty
\bibitem [{\citenamefont {Hirose}\ \emph {et~al.}(2021)\citenamefont {Hirose},
  \citenamefont {Wood},\ and\ \citenamefont {Vočadlo}}]{hirose21}%
  \BibitemOpen
  \bibfield  {author} {\bibinfo {author} {\bibfnamefont {K.}~\bibnamefont
  {Hirose}}, \bibinfo {author} {\bibfnamefont {B.}~\bibnamefont {Wood}},\ and\
  \bibinfo {author} {\bibfnamefont {L.}~\bibnamefont {Vočadlo}},\ }\bibfield
  {title} {\bibinfo {title} {Light elements in the {E}arth’s core},\ }\href
  {https://doi.org/https://doi.org/10.1038/s43017-021-00203-6} {\bibfield
  {journal} {\bibinfo  {journal} {Nature Reviews Earth and Environment}\
  }\textbf {\bibinfo {volume} {2}},\ \bibinfo {pages} {645–658} (\bibinfo
  {year} {2021})}\BibitemShut {NoStop}%
\bibitem [{\citenamefont {Ozawa}\ \emph {et~al.}(2011)\citenamefont {Ozawa},
  \citenamefont {Takahashi}, \citenamefont {Hirose}, \citenamefont {Ohishi},\
  and\ \citenamefont {Hirao}}]{ozawa}%
  \BibitemOpen
  \bibfield  {author} {\bibinfo {author} {\bibfnamefont {H.}~\bibnamefont
  {Ozawa}}, \bibinfo {author} {\bibfnamefont {F.}~\bibnamefont {Takahashi}},
  \bibinfo {author} {\bibfnamefont {K.}~\bibnamefont {Hirose}}, \bibinfo
  {author} {\bibfnamefont {Y.}~\bibnamefont {Ohishi}},\ and\ \bibinfo {author}
  {\bibfnamefont {N.}~\bibnamefont {Hirao}},\ }\bibfield  {title} {\bibinfo
  {title} {Phase {T}ransition of {F}e{O} and {S}tratification in {E}arth’s
  {O}uter {C}ore},\ }\href
  {https://doi.org/https://doi.org/10.1126/science.1208265} {\bibfield
  {journal} {\bibinfo  {journal} {Science}\ }\textbf {\bibinfo {volume}
  {334}},\ \bibinfo {pages} {792} (\bibinfo {year} {2011})}\BibitemShut
  {NoStop}%
\bibitem [{\citenamefont {Lavina}\ \emph {et~al.}(2011)\citenamefont {Lavina},
  \citenamefont {Dera}, \citenamefont {Kim}, \citenamefont {Meng},
  \citenamefont {Downs}, \citenamefont {Weck}, \citenamefont {Sutton},\ and\
  \citenamefont {Zhao}}]{lavina}%
  \BibitemOpen
  \bibfield  {author} {\bibinfo {author} {\bibfnamefont {B.}~\bibnamefont
  {Lavina}}, \bibinfo {author} {\bibfnamefont {P.}~\bibnamefont {Dera}},
  \bibinfo {author} {\bibfnamefont {E.}~\bibnamefont {Kim}}, \bibinfo {author}
  {\bibfnamefont {Y.}~\bibnamefont {Meng}}, \bibinfo {author} {\bibfnamefont
  {R.~T.}\ \bibnamefont {Downs}}, \bibinfo {author} {\bibfnamefont {P.~F.}\
  \bibnamefont {Weck}}, \bibinfo {author} {\bibfnamefont {S.~R.}\ \bibnamefont
  {Sutton}},\ and\ \bibinfo {author} {\bibfnamefont {Y.}~\bibnamefont {Zhao}},\
  }\bibfield  {title} {\bibinfo {title} {Discovery of the recoverable
  high-pressure iron oxide \ch{Fe4O5}},\ }\href
  {https://doi.org/https://doi.org/10.1073/pnas.110757310} {\bibfield
  {journal} {\bibinfo  {journal} {Proceedings of the National Academy of
  Sciences}\ }\textbf {\bibinfo {volume} {108}},\ \bibinfo {pages} {17281}
  (\bibinfo {year} {2011})}\BibitemShut {NoStop}%
\bibitem [{\citenamefont {Lavina}\ and\ \citenamefont
  {Meng}(2015)}]{lavina2015}%
  \BibitemOpen
  \bibfield  {author} {\bibinfo {author} {\bibfnamefont {B.}~\bibnamefont
  {Lavina}}\ and\ \bibinfo {author} {\bibfnamefont {Y.}~\bibnamefont {Meng}},\
  }\bibfield  {title} {\bibinfo {title} {Discovery of the recoverable
  high-pressure iron oxide \ch{Fe4O5}},\ }\href
  {https://doi.org/https://doi.org/10.1126/sciadv.140026} {\bibfield  {journal}
  {\bibinfo  {journal} {Science Advances}\ }\textbf {\bibinfo {volume} {1}},\
  \bibinfo {pages} {e1400260} (\bibinfo {year} {2015})}\BibitemShut {NoStop}%
\bibitem [{\citenamefont {Sinmyo}\ \emph {et~al.}(2016)\citenamefont {Sinmyo},
  \citenamefont {Bykova}, \citenamefont {Ovsyannikov}, \citenamefont
  {McCammon}, \citenamefont {Kupenko}, \citenamefont {Ismailova},\ and\
  \citenamefont {Dubrovinsky}}]{Sinmyo}%
  \BibitemOpen
  \bibfield  {author} {\bibinfo {author} {\bibfnamefont {R.}~\bibnamefont
  {Sinmyo}}, \bibinfo {author} {\bibfnamefont {E.}~\bibnamefont {Bykova}},
  \bibinfo {author} {\bibfnamefont {S.~V.}\ \bibnamefont {Ovsyannikov}},
  \bibinfo {author} {\bibfnamefont {C.}~\bibnamefont {McCammon}}, \bibinfo
  {author} {\bibfnamefont {I.}~\bibnamefont {Kupenko}}, \bibinfo {author}
  {\bibfnamefont {L.}~\bibnamefont {Ismailova}},\ and\ \bibinfo {author}
  {\bibfnamefont {L.}~\bibnamefont {Dubrovinsky}},\ }\bibfield  {title}
  {\bibinfo {title} {Discovery of \ch{Fe7O9}: a new iron oxide with a complex
  monoclinic structure},\ }\href
  {https://doi.org/https://doi.org/10.1038/srep32852} {\bibfield  {journal}
  {\bibinfo  {journal} {Scientific Reports}\ }\textbf {\bibinfo {volume} {6}},\
  \bibinfo {pages} {6622} (\bibinfo {year} {2016})}\BibitemShut {NoStop}%
\bibitem [{\citenamefont {Pasternak}\ \emph {et~al.}(1999)\citenamefont
  {Pasternak}, \citenamefont {Rozenberg}, \citenamefont {Machavariani},
  \citenamefont {Naaman}, \citenamefont {Taylor},\ and\ \citenamefont
  {Jeanloz}}]{pasternak}%
  \BibitemOpen
  \bibfield  {author} {\bibinfo {author} {\bibfnamefont {M.~P.}\ \bibnamefont
  {Pasternak}}, \bibinfo {author} {\bibfnamefont {G.~K.}\ \bibnamefont
  {Rozenberg}}, \bibinfo {author} {\bibfnamefont {G.~Y.}\ \bibnamefont
  {Machavariani}}, \bibinfo {author} {\bibfnamefont {O.}~\bibnamefont
  {Naaman}}, \bibinfo {author} {\bibfnamefont {R.~D.}\ \bibnamefont {Taylor}},\
  and\ \bibinfo {author} {\bibfnamefont {R.}~\bibnamefont {Jeanloz}},\
  }\bibfield  {title} {\bibinfo {title} {Breakdown of the {M}ott-{H}ubbard
  {S}tate in \ch{Fe2O3}: A {F}irst-{O}rder {I}nsulator-{M}etal {T}ransition
  with {C}ollapse of {M}agnetism at 50 {GP}a},\ }\href
  {https://doi.org/https://doi.org/10.1103/PhysRevLett.82.4663} {\bibfield
  {journal} {\bibinfo  {journal} {Physical Review Letters}\ }\textbf {\bibinfo
  {volume} {82}},\ \bibinfo {pages} {4663} (\bibinfo {year}
  {1999})}\BibitemShut {NoStop}%
\bibitem [{\citenamefont {Greenberg}\ \emph {et~al.}(2018)\citenamefont
  {Greenberg}, \citenamefont {Leonov}, \citenamefont {Layek}, \citenamefont
  {Konopkova}, \citenamefont {Pasternak}, \citenamefont {Dubrovinsky},
  \citenamefont {Jeanloz}, \citenamefont {Abrikosov},\ and\ \citenamefont
  {Rozenberg}}]{greenberg}%
  \BibitemOpen
  \bibfield  {author} {\bibinfo {author} {\bibfnamefont {E.}~\bibnamefont
  {Greenberg}}, \bibinfo {author} {\bibfnamefont {I.}~\bibnamefont {Leonov}},
  \bibinfo {author} {\bibfnamefont {S.}~\bibnamefont {Layek}}, \bibinfo
  {author} {\bibfnamefont {Z.}~\bibnamefont {Konopkova}}, \bibinfo {author}
  {\bibfnamefont {M.~P.}\ \bibnamefont {Pasternak}}, \bibinfo {author}
  {\bibfnamefont {L.}~\bibnamefont {Dubrovinsky}}, \bibinfo {author}
  {\bibfnamefont {R.}~\bibnamefont {Jeanloz}}, \bibinfo {author} {\bibfnamefont
  {I.~A.}\ \bibnamefont {Abrikosov}},\ and\ \bibinfo {author} {\bibfnamefont
  {G.~K.}\ \bibnamefont {Rozenberg}},\ }\bibfield  {title} {\bibinfo {title}
  {Pressure-{I}nduced {S}ite-{S}elective {M}ott {I}nsulator-{M}etal transition
  in \ch{Fe2O3}},\ }\href
  {https://doi.org/https://doi.org/10.1103/PhysRevX.8.031059} {\bibfield
  {journal} {\bibinfo  {journal} {Physical Review X}\ }\textbf {\bibinfo
  {volume} {8}},\ \bibinfo {pages} {031059} (\bibinfo {year}
  {2018})}\BibitemShut {NoStop}%
\bibitem [{\citenamefont {Badro}\ \emph {et~al.}(2002)\citenamefont {Badro},
  \citenamefont {Fiquet}, \citenamefont {Struzhkin}, \citenamefont
  {Somayazulu}, \citenamefont {Mao}, \citenamefont {Shen},\ and\ \citenamefont
  {LeBihan}}]{badro}%
  \BibitemOpen
  \bibfield  {author} {\bibinfo {author} {\bibfnamefont {J.}~\bibnamefont
  {Badro}}, \bibinfo {author} {\bibfnamefont {G.}~\bibnamefont {Fiquet}},
  \bibinfo {author} {\bibfnamefont {V.~V.}\ \bibnamefont {Struzhkin}}, \bibinfo
  {author} {\bibfnamefont {M.}~\bibnamefont {Somayazulu}}, \bibinfo {author}
  {\bibfnamefont {H.-K.}\ \bibnamefont {Mao}}, \bibinfo {author} {\bibfnamefont
  {G.}~\bibnamefont {Shen}},\ and\ \bibinfo {author} {\bibfnamefont
  {T.}~\bibnamefont {LeBihan}},\ }\bibfield  {title} {\bibinfo {title} {{Nature
  of the high-pressure transition in \ch{Fe2O3} hematite}},\ }\href
  {https://doi.org/10.1103/PhysRevLett.89.205504} {\bibfield  {journal}
  {\bibinfo  {journal} {Physical Review Letters}\ }\textbf {\bibinfo {volume}
  {89}},\ \bibinfo {pages} {205504} (\bibinfo {year} {2002})}\BibitemShut
  {NoStop}%
\bibitem [{\citenamefont {Sanson}\ \emph {et~al.}(2016)\citenamefont {Sanson},
  \citenamefont {Kantor}, \citenamefont {Cerantola}, \citenamefont {Irifune},
  \citenamefont {Carnera},\ and\ \citenamefont {Pascarelli}}]{sanson}%
  \BibitemOpen
  \bibfield  {author} {\bibinfo {author} {\bibfnamefont {A.}~\bibnamefont
  {Sanson}}, \bibinfo {author} {\bibfnamefont {I.}~\bibnamefont {Kantor}},
  \bibinfo {author} {\bibfnamefont {V.}~\bibnamefont {Cerantola}}, \bibinfo
  {author} {\bibfnamefont {T.}~\bibnamefont {Irifune}}, \bibinfo {author}
  {\bibfnamefont {A.}~\bibnamefont {Carnera}},\ and\ \bibinfo {author}
  {\bibfnamefont {S.}~\bibnamefont {Pascarelli}},\ }\bibfield  {title}
  {\bibinfo {title} {{Local Structure and Spin Transition in \ch{Fe2O3}
  Hematite at High-Pressure}},\ }\href
  {https://doi.org/10.1103/PhysRevB.94.014112} {\bibfield  {journal} {\bibinfo
  {journal} {Physical Review B}\ }\textbf {\bibinfo {volume} {94}},\ \bibinfo
  {pages} {014112} (\bibinfo {year} {2016})}\BibitemShut {NoStop}%
\bibitem [{\citenamefont {Bykova}\ \emph {et~al.}(2016)\citenamefont {Bykova},
  \citenamefont {Dubrovinsky}, \citenamefont {Dubrovinskaia}, \citenamefont
  {Bykov}, \citenamefont {McCammon}, \citenamefont {Ovsyannikov}, \citenamefont
  {Liermann}, \citenamefont {Kupenko}, \citenamefont {Chumakov}, \citenamefont
  {Ruffer}, \citenamefont {Hanfland},\ and\ \citenamefont
  {Prakapenka}}]{bykova}%
  \BibitemOpen
  \bibfield  {author} {\bibinfo {author} {\bibfnamefont {E.}~\bibnamefont
  {Bykova}}, \bibinfo {author} {\bibfnamefont {L.}~\bibnamefont {Dubrovinsky}},
  \bibinfo {author} {\bibfnamefont {N.}~\bibnamefont {Dubrovinskaia}}, \bibinfo
  {author} {\bibfnamefont {M.}~\bibnamefont {Bykov}}, \bibinfo {author}
  {\bibfnamefont {C.}~\bibnamefont {McCammon}}, \bibinfo {author}
  {\bibfnamefont {S.}~\bibnamefont {Ovsyannikov}}, \bibinfo {author}
  {\bibfnamefont {H.-P.}\ \bibnamefont {Liermann}}, \bibinfo {author}
  {\bibfnamefont {I.}~\bibnamefont {Kupenko}}, \bibinfo {author} {\bibfnamefont
  {A.}~\bibnamefont {Chumakov}}, \bibinfo {author} {\bibfnamefont
  {R.}~\bibnamefont {Ruffer}}, \bibinfo {author} {\bibfnamefont
  {M.}~\bibnamefont {Hanfland}},\ and\ \bibinfo {author} {\bibfnamefont
  {V.}~\bibnamefont {Prakapenka}},\ }\bibfield  {title} {\bibinfo {title}
  {Structural complexity of simple \ch{Fe2O3} at high pressures and
  temperatures},\ }\href {https://doi.org/https://doi.org/10.1038/ncomms10661}
  {\bibfield  {journal} {\bibinfo  {journal} {Nature communications}\ }\textbf
  {\bibinfo {volume} {7}},\ \bibinfo {pages} {10661} (\bibinfo {year}
  {2016})}\BibitemShut {NoStop}%
\bibitem [{\citenamefont {Amouretti}\ \emph
  {et~al.}(2024{\natexlab{a}})\citenamefont {Amouretti}, \citenamefont
  {Crepisson}, \citenamefont {Azadi}, \citenamefont {Cabaret}, \citenamefont
  {Campbell}, \citenamefont {Chin}, \citenamefont {Colin}, \citenamefont
  {Collins}, \citenamefont {Crandall}, \citenamefont {Fiquet}, \citenamefont
  {Forte}, \citenamefont {Gawne}, \citenamefont {Guyot}, \citenamefont
  {Heighway}, \citenamefont {Lee}, \citenamefont {McGonegle}, \citenamefont
  {Nagler}, \citenamefont {Pintor}, \citenamefont {Polsin}, \citenamefont
  {Rousse}, \citenamefont {Shi}, \citenamefont {Smith}, \citenamefont {Wark},
  \citenamefont {Vinko},\ and\ \citenamefont {Harmand}}]{Alexis}%
  \BibitemOpen
  \bibfield  {author} {\bibinfo {author} {\bibfnamefont {A.}~\bibnamefont
  {Amouretti}}, \bibinfo {author} {\bibfnamefont {C.}~\bibnamefont
  {Crepisson}}, \bibinfo {author} {\bibfnamefont {S.}~\bibnamefont {Azadi}},
  \bibinfo {author} {\bibfnamefont {D.}~\bibnamefont {Cabaret}}, \bibinfo
  {author} {\bibfnamefont {T.}~\bibnamefont {Campbell}}, \bibinfo {author}
  {\bibfnamefont {D.~A.}\ \bibnamefont {Chin}}, \bibinfo {author}
  {\bibfnamefont {B.}~\bibnamefont {Colin}}, \bibinfo {author} {\bibfnamefont
  {G.~R.}\ \bibnamefont {Collins}}, \bibinfo {author} {\bibfnamefont
  {L.}~\bibnamefont {Crandall}}, \bibinfo {author} {\bibfnamefont
  {G.}~\bibnamefont {Fiquet}}, \bibinfo {author} {\bibfnamefont
  {A.}~\bibnamefont {Forte}}, \bibinfo {author} {\bibfnamefont
  {T.}~\bibnamefont {Gawne}}, \bibinfo {author} {\bibfnamefont
  {F.}~\bibnamefont {Guyot}}, \bibinfo {author} {\bibfnamefont
  {P.}~\bibnamefont {Heighway}}, \bibinfo {author} {\bibfnamefont
  {H.}~\bibnamefont {Lee}}, \bibinfo {author} {\bibfnamefont {D.}~\bibnamefont
  {McGonegle}}, \bibinfo {author} {\bibfnamefont {B.}~\bibnamefont {Nagler}},
  \bibinfo {author} {\bibfnamefont {J.}~\bibnamefont {Pintor}}, \bibinfo
  {author} {\bibfnamefont {D.}~\bibnamefont {Polsin}}, \bibinfo {author}
  {\bibfnamefont {G.}~\bibnamefont {Rousse}}, \bibinfo {author} {\bibfnamefont
  {Y.}~\bibnamefont {Shi}}, \bibinfo {author} {\bibfnamefont {E.}~\bibnamefont
  {Smith}}, \bibinfo {author} {\bibfnamefont {J.~S.}\ \bibnamefont {Wark}},
  \bibinfo {author} {\bibfnamefont {S.~M.}\ \bibnamefont {Vinko}},\ and\
  \bibinfo {author} {\bibfnamefont {M.}~\bibnamefont {Harmand}},\ }\bibfield
  {title} {\bibinfo {title} {Phase transitions of \ch{Fe2O3} under laser shock
  compression}\ }\href {https://doi.org/10.48550/arXiv.2402.18432}
  {10.48550/arXiv.2402.18432} (\bibinfo {year}
  {2024}{\natexlab{a}})\BibitemShut {NoStop}%
\bibitem [{\citenamefont {Nagler}\ \emph {et~al.}(2015)\citenamefont {Nagler},
  \citenamefont {Arnold}, \citenamefont {Bouchard}, \citenamefont {Boyce},
  \citenamefont {Boyce}, \citenamefont {Callen}, \citenamefont {Campell},
  \citenamefont {Curiel}, \citenamefont {Galtier}, \citenamefont {Garofoli},
  \citenamefont {Granados}, \citenamefont {Hastings}, \citenamefont {Hays},
  \citenamefont {Heimann}, \citenamefont {Lee}, \citenamefont {Milathianaki},
  \citenamefont {Plummer}, \citenamefont {Schropp}, \citenamefont {Wallace},
  \citenamefont {Welch}, \citenamefont {White}, \citenamefont {Xing},
  \citenamefont {Yin}, \citenamefont {Young}, \citenamefont {Zastrau},\ and\
  \citenamefont {Lee}}]{mec}%
  \BibitemOpen
  \bibfield  {author} {\bibinfo {author} {\bibfnamefont {B.}~\bibnamefont
  {Nagler}}, \bibinfo {author} {\bibfnamefont {B.}~\bibnamefont {Arnold}},
  \bibinfo {author} {\bibfnamefont {G.}~\bibnamefont {Bouchard}}, \bibinfo
  {author} {\bibfnamefont {R.~F.}\ \bibnamefont {Boyce}}, \bibinfo {author}
  {\bibfnamefont {R.~M.}\ \bibnamefont {Boyce}}, \bibinfo {author}
  {\bibfnamefont {A.}~\bibnamefont {Callen}}, \bibinfo {author} {\bibfnamefont
  {M.}~\bibnamefont {Campell}}, \bibinfo {author} {\bibfnamefont
  {R.}~\bibnamefont {Curiel}}, \bibinfo {author} {\bibfnamefont
  {E.}~\bibnamefont {Galtier}}, \bibinfo {author} {\bibfnamefont
  {J.}~\bibnamefont {Garofoli}}, \bibinfo {author} {\bibfnamefont
  {E.}~\bibnamefont {Granados}}, \bibinfo {author} {\bibfnamefont
  {J.}~\bibnamefont {Hastings}}, \bibinfo {author} {\bibfnamefont
  {G.}~\bibnamefont {Hays}}, \bibinfo {author} {\bibfnamefont {P.}~\bibnamefont
  {Heimann}}, \bibinfo {author} {\bibfnamefont {R.~W.}\ \bibnamefont {Lee}},
  \bibinfo {author} {\bibfnamefont {D.}~\bibnamefont {Milathianaki}}, \bibinfo
  {author} {\bibfnamefont {L.}~\bibnamefont {Plummer}}, \bibinfo {author}
  {\bibfnamefont {A.}~\bibnamefont {Schropp}}, \bibinfo {author} {\bibfnamefont
  {A.}~\bibnamefont {Wallace}}, \bibinfo {author} {\bibfnamefont
  {M.}~\bibnamefont {Welch}}, \bibinfo {author} {\bibfnamefont
  {W.}~\bibnamefont {White}}, \bibinfo {author} {\bibfnamefont
  {Z.}~\bibnamefont {Xing}}, \bibinfo {author} {\bibfnamefont {J.}~\bibnamefont
  {Yin}}, \bibinfo {author} {\bibfnamefont {J.}~\bibnamefont {Young}}, \bibinfo
  {author} {\bibfnamefont {U.}~\bibnamefont {Zastrau}},\ and\ \bibinfo {author}
  {\bibfnamefont {H.~J.}\ \bibnamefont {Lee}},\ }\bibfield  {title} {\bibinfo
  {title} {The {M}atter in {E}xtreme {C}onditions instrument at the {L}inac
  {C}oherent {L}ight {S}ource},\ }\href
  {https://doi.org/https://doi.org/10.1107/S1600577515004865} {\bibfield
  {journal} {\bibinfo  {journal} {Journal of Synchrotron Radiation}\ }\textbf
  {\bibinfo {volume} {22}},\ \bibinfo {pages} {520} (\bibinfo {year}
  {2015})}\BibitemShut {NoStop}%
\bibitem [{\citenamefont {Yabashi}\ \emph {et~al.}(2015)\citenamefont
  {Yabashi}, \citenamefont {Tanaka},\ and\ \citenamefont {Ishikawa}}]{sacla}%
  \BibitemOpen
  \bibfield  {author} {\bibinfo {author} {\bibfnamefont {M.}~\bibnamefont
  {Yabashi}}, \bibinfo {author} {\bibfnamefont {H.}~\bibnamefont {Tanaka}},\
  and\ \bibinfo {author} {\bibfnamefont {T.}~\bibnamefont {Ishikawa}},\
  }\bibfield  {title} {\bibinfo {title} {Overview of the {SACLA} facility},\
  }\href {https://doi.org/https://doi:10.1107/S1600577515004658} {\bibfield
  {journal} {\bibinfo  {journal} {Journal of Synchrotron Radiation}\ }\textbf
  {\bibinfo {volume} {22(pt 3)}},\ \bibinfo {pages} {477–484} (\bibinfo
  {year} {2015})}\BibitemShut {NoStop}%
\bibitem [{\citenamefont {Prescher}\ and\ \citenamefont
  {Prakapenka}(2015)}]{presher}%
  \BibitemOpen
  \bibfield  {author} {\bibinfo {author} {\bibfnamefont {C.}~\bibnamefont
  {Prescher}}\ and\ \bibinfo {author} {\bibfnamefont {V.~B.}\ \bibnamefont
  {Prakapenka}},\ }\bibfield  {title} {\bibinfo {title} {Dioptas: a program for
  reduction of two-dimensional x-ray diffraction data and data exploration},\
  }\href {https://doi.org/https://doi.org/10.1080/08957959.2015.1059835}
  {\bibfield  {journal} {\bibinfo  {journal} {High Pressure Research}\ }\textbf
  {\bibinfo {volume} {35}},\ \bibinfo {pages} {223} (\bibinfo {year}
  {2015})}\BibitemShut {NoStop}%
\bibitem [{\citenamefont {Falk}(2018)}]{falk}%
  \BibitemOpen
  \bibfield  {author} {\bibinfo {author} {\bibfnamefont {K.}~\bibnamefont
  {Falk}},\ }\bibfield  {title} {\bibinfo {title} {Experimental methods for
  warm dense matter research},\ }\href
  {https://doi.org/https://doi.org/10.1017/hpl.2018.53} {\bibfield  {journal}
  {\bibinfo  {journal} {High Power Laser Science and Engineering}\ }\textbf
  {\bibinfo {volume} {6}},\ \bibinfo {pages} {e59} (\bibinfo {year}
  {2018})}\BibitemShut {NoStop}%
\bibitem [{\citenamefont {Blake}\ \emph {et~al.}(1966)\citenamefont {Blake},
  \citenamefont {Hessevick}, \citenamefont {Zoltai},\ and\ \citenamefont
  {Finger}}]{blake}%
  \BibitemOpen
  \bibfield  {author} {\bibinfo {author} {\bibfnamefont {R.~L.}\ \bibnamefont
  {Blake}}, \bibinfo {author} {\bibfnamefont {R.~E.}\ \bibnamefont
  {Hessevick}}, \bibinfo {author} {\bibfnamefont {T.}~\bibnamefont {Zoltai}},\
  and\ \bibinfo {author} {\bibfnamefont {L.~W.}\ \bibnamefont {Finger}},\
  }\bibfield  {title} {\bibinfo {title} {Refinement of the {H}ematite
  {S}tructure},\ }\href@noop {} {\bibfield  {journal} {\bibinfo  {journal}
  {American Mineralogist}\ }\textbf {\bibinfo {volume} {51}},\ \bibinfo {pages}
  {123} (\bibinfo {year} {1966})}\BibitemShut {NoStop}%
\bibitem [{\citenamefont {Barker}\ and\ \citenamefont
  {Hollenbach}(1972)}]{barker}%
  \BibitemOpen
  \bibfield  {author} {\bibinfo {author} {\bibfnamefont {L.~M.}\ \bibnamefont
  {Barker}}\ and\ \bibinfo {author} {\bibfnamefont {R.~E.}\ \bibnamefont
  {Hollenbach}},\ }\bibfield  {title} {\bibinfo {title} {Laser interferometer
  for measuring high velocities of any reflecting surface},\ }\href
  {https://doi.org/https://doi.org/10.1063/1.1660986} {\bibfield  {journal}
  {\bibinfo  {journal} {Journal of Applied Physics}\ }\textbf {\bibinfo
  {volume} {43}},\ \bibinfo {pages} {4669} (\bibinfo {year}
  {1972})}\BibitemShut {NoStop}%
\bibitem [{\citenamefont {Cao}\ \emph {et~al.}(2017)\citenamefont {Cao},
  \citenamefont {Wang}, \citenamefont {Li}, \citenamefont {Xu}, \citenamefont
  {Liu}, \citenamefont {Yu}, \citenamefont {Qin}, \citenamefont {Zhu},
  \citenamefont {Tang}, \citenamefont {He}, \citenamefont {Meng}, \citenamefont
  {Zhang},\ and\ \citenamefont {Peng}}]{cao}%
  \BibitemOpen
  \bibfield  {author} {\bibinfo {author} {\bibfnamefont {X.}~\bibnamefont
  {Cao}}, \bibinfo {author} {\bibfnamefont {Y.}~\bibnamefont {Wang}}, \bibinfo
  {author} {\bibfnamefont {X.}~\bibnamefont {Li}}, \bibinfo {author}
  {\bibfnamefont {L.}~\bibnamefont {Xu}}, \bibinfo {author} {\bibfnamefont
  {L.}~\bibnamefont {Liu}}, \bibinfo {author} {\bibfnamefont {Y.}~\bibnamefont
  {Yu}}, \bibinfo {author} {\bibfnamefont {R.}~\bibnamefont {Qin}}, \bibinfo
  {author} {\bibfnamefont {W.}~\bibnamefont {Zhu}}, \bibinfo {author}
  {\bibfnamefont {S.}~\bibnamefont {Tang}}, \bibinfo {author} {\bibfnamefont
  {L.}~\bibnamefont {He}}, \bibinfo {author} {\bibfnamefont {C.}~\bibnamefont
  {Meng}}, \bibinfo {author} {\bibfnamefont {B.}~\bibnamefont {Zhang}},\ and\
  \bibinfo {author} {\bibfnamefont {X.}~\bibnamefont {Peng}},\ }\bibfield
  {title} {\bibinfo {title} {Refractive index and phase transformation of
  sapphire under shock pressures up to 210 {G}{P}a},\ }\href
  {https://doi.org/10.1063/1.4978746} {\bibfield  {journal} {\bibinfo
  {journal} {Journal of Applied physics}\ }\textbf {\bibinfo {volume} {121}},\
  \bibinfo {pages} {115903} (\bibinfo {year} {2017})}\BibitemShut {NoStop}%
\bibitem [{\citenamefont {Ramis}\ \emph {et~al.}(2009)\citenamefont {Ramis},
  \citenamefont {ter Vehn},\ and\ \citenamefont {Ramírez}}]{ramis}%
  \BibitemOpen
  \bibfield  {author} {\bibinfo {author} {\bibfnamefont {R.}~\bibnamefont
  {Ramis}}, \bibinfo {author} {\bibfnamefont {J.~M.}\ \bibnamefont {ter
  Vehn}},\ and\ \bibinfo {author} {\bibfnamefont {J.}~\bibnamefont
  {Ramírez}},\ }\bibfield  {title} {\bibinfo {title} {Multi2d – a computer
  code for two-dimensional radiation hydrodynamics},\ }\href
  {https://doi.org/https://doi.org/10.1016/j.cpc.2008.12.033} {\bibfield
  {journal} {\bibinfo  {journal} {Computer Physics Communications}\ }\textbf
  {\bibinfo {volume} {180}},\ \bibinfo {pages} {977} (\bibinfo {year}
  {2009})}\BibitemShut {NoStop}%
\bibitem [{\citenamefont {Hadju}(1972)}]{hadju}%
  \BibitemOpen
  \bibfield  {author} {\bibinfo {author} {\bibfnamefont {F.}~\bibnamefont
  {Hadju}},\ }\bibfield  {title} {\bibinfo {title} {Revised parameters of the
  analytic fits for coherent and incoherent scattered x-ray intensities of the
  first 36 atoms},\ }\href
  {https://doi.org/https://doi.org/10.1107/S0567739472000671} {\bibfield
  {journal} {\bibinfo  {journal} {Acta Crystallographica Section A}\ }\textbf
  {\bibinfo {volume} {A28}},\ \bibinfo {pages} {250} (\bibinfo {year}
  {1972})}\BibitemShut {NoStop}%
\bibitem [{\citenamefont {Heinen}\ and\ \citenamefont
  {Drewitt}(2022)}]{liquiddiffract}%
  \BibitemOpen
  \bibfield  {author} {\bibinfo {author} {\bibfnamefont {B.~J.}\ \bibnamefont
  {Heinen}}\ and\ \bibinfo {author} {\bibfnamefont {J.~W.~E.}\ \bibnamefont
  {Drewitt}},\ }\bibfield  {title} {\bibinfo {title} {Liquiddiffract: Software
  for liquid total scattering analysis},\ }\href
  {https://doi.org/https://doi.org/10.1007/s00269-022-01186-6} {\bibfield
  {journal} {\bibinfo  {journal} {Physics and Chemistry of Minerals}\ }\textbf
  {\bibinfo {volume} {49}},\ \bibinfo {pages} {1432} (\bibinfo {year}
  {2022})}\BibitemShut {NoStop}%
\bibitem [{\citenamefont {Eggert}\ \emph {et~al.}(2002)\citenamefont {Eggert},
  \citenamefont {Weck}, \citenamefont {Loubeyre},\ and\ \citenamefont
  {Mezouar}}]{eggert}%
  \BibitemOpen
  \bibfield  {author} {\bibinfo {author} {\bibfnamefont {J.~H.}\ \bibnamefont
  {Eggert}}, \bibinfo {author} {\bibfnamefont {G.}~\bibnamefont {Weck}},
  \bibinfo {author} {\bibfnamefont {P.}~\bibnamefont {Loubeyre}},\ and\
  \bibinfo {author} {\bibfnamefont {M.}~\bibnamefont {Mezouar}},\ }\bibfield
  {title} {\bibinfo {title} {Quantitative structure factor and density
  measurements of high-pressure fluids in diamond anvil cells by x-ray
  diffraction: Argon and water},\ }\href
  {https://doi.org/https://doi.org/10.1103/PhysRevB.65.174105} {\bibfield
  {journal} {\bibinfo  {journal} {Physical Review B}\ }\textbf {\bibinfo
  {volume} {65}},\ \bibinfo {pages} {174105} (\bibinfo {year}
  {2002})}\BibitemShut {NoStop}%
\bibitem [{\citenamefont {Sanloup}\ \emph {et~al.}(2000)\citenamefont
  {Sanloup}, \citenamefont {Guyot}, \citenamefont {Gillet}, \citenamefont
  {Fiquet}, \citenamefont {Hemley}, \citenamefont {Mezouar},\ and\
  \citenamefont {Martinez}}]{sanloup}%
  \BibitemOpen
  \bibfield  {author} {\bibinfo {author} {\bibfnamefont {C.}~\bibnamefont
  {Sanloup}}, \bibinfo {author} {\bibfnamefont {F.}~\bibnamefont {Guyot}},
  \bibinfo {author} {\bibfnamefont {P.}~\bibnamefont {Gillet}}, \bibinfo
  {author} {\bibfnamefont {G.}~\bibnamefont {Fiquet}}, \bibinfo {author}
  {\bibfnamefont {R.~J.}\ \bibnamefont {Hemley}}, \bibinfo {author}
  {\bibfnamefont {M.}~\bibnamefont {Mezouar}},\ and\ \bibinfo {author}
  {\bibfnamefont {I.}~\bibnamefont {Martinez}},\ }\bibfield  {title} {\bibinfo
  {title} {Structural changes in liquid {F}e at high pressures and high
  temperatures from synchrotron x-ray diffraction},\ }\href
  {https://doi.org/https://doi.org/10.1209/epl/i2000-00417-3} {\bibfield
  {journal} {\bibinfo  {journal} {Europhysics Letters}\ }\textbf {\bibinfo
  {volume} {52}},\ \bibinfo {pages} {151} (\bibinfo {year} {2000})}\BibitemShut
  {NoStop}%
\bibitem [{\citenamefont {Hohenberg}\ and\ \citenamefont
  {Kohn}(1964)}]{Hohenberg}%
  \BibitemOpen
  \bibfield  {author} {\bibinfo {author} {\bibfnamefont {P.}~\bibnamefont
  {Hohenberg}}\ and\ \bibinfo {author} {\bibfnamefont {W.}~\bibnamefont
  {Kohn}},\ }\bibfield  {title} {\bibinfo {title} {Inhomogeneous electron
  gas},\ }\href@noop {} {\bibfield  {journal} {\bibinfo  {journal} {Physical
  Review}\ }\textbf {\bibinfo {volume} {136}},\ \bibinfo {pages} {B864}
  (\bibinfo {year} {1964})}\BibitemShut {NoStop}%
\bibitem [{\citenamefont {Kohn}\ and\ \citenamefont {Sham}(1965)}]{Kohn}%
  \BibitemOpen
  \bibfield  {author} {\bibinfo {author} {\bibfnamefont {W.}~\bibnamefont
  {Kohn}}\ and\ \bibinfo {author} {\bibfnamefont {L.~J.}\ \bibnamefont
  {Sham}},\ }\bibfield  {title} {\bibinfo {title} {Self-consistent equations
  including exchange and correlation effects},\ }\href@noop {} {\bibfield
  {journal} {\bibinfo  {journal} {Physical Review}\ }\textbf {\bibinfo {volume}
  {140}},\ \bibinfo {pages} {A1133} (\bibinfo {year} {1965})}\BibitemShut
  {NoStop}%
\bibitem [{\citenamefont {Anisimov}\ \emph {et~al.}(1991)\citenamefont
  {Anisimov}, \citenamefont {Zaanen},\ and\ \citenamefont
  {Andersen}}]{Anisimov}%
  \BibitemOpen
  \bibfield  {author} {\bibinfo {author} {\bibfnamefont {V.~I.}\ \bibnamefont
  {Anisimov}}, \bibinfo {author} {\bibfnamefont {J.}~\bibnamefont {Zaanen}},\
  and\ \bibinfo {author} {\bibfnamefont {O.~K.}\ \bibnamefont {Andersen}},\
  }\bibfield  {title} {\bibinfo {title} {Band theory and mott insulators:
  Hubbard {U} instead of stoner {I}},\ }\href@noop {} {\bibfield  {journal}
  {\bibinfo  {journal} {Physical Review B}\ }\textbf {\bibinfo {volume} {44}},\
  \bibinfo {pages} {943} (\bibinfo {year} {1991})}\BibitemShut {NoStop}%
\bibitem [{\citenamefont {Liechtenstein}\ \emph {et~al.}(1995)\citenamefont
  {Liechtenstein}, \citenamefont {Anisimov},\ and\ \citenamefont
  {Zaanen}}]{Liechtenstein}%
  \BibitemOpen
  \bibfield  {author} {\bibinfo {author} {\bibfnamefont {A.~I.}\ \bibnamefont
  {Liechtenstein}}, \bibinfo {author} {\bibfnamefont {V.~I.}\ \bibnamefont
  {Anisimov}},\ and\ \bibinfo {author} {\bibfnamefont {J.}~\bibnamefont
  {Zaanen}},\ }\bibfield  {title} {\bibinfo {title} {Density-functional theory
  and strong interactions: Orbital ordering in {M}ott-{H}ubbard insulators},\
  }\href@noop {} {\bibfield  {journal} {\bibinfo  {journal} {Physical Review
  B}\ }\textbf {\bibinfo {volume} {52}},\ \bibinfo {pages} {R5467} (\bibinfo
  {year} {1995})}\BibitemShut {NoStop}%
\bibitem [{\citenamefont {Dudarev}\ \emph {et~al.}(1998)\citenamefont
  {Dudarev}, \citenamefont {Botton}, \citenamefont {Savrasov}, \citenamefont
  {Humphreys},\ and\ \citenamefont {Sutton}}]{Dudarev}%
  \BibitemOpen
  \bibfield  {author} {\bibinfo {author} {\bibfnamefont {S.~L.}\ \bibnamefont
  {Dudarev}}, \bibinfo {author} {\bibfnamefont {G.~A.}\ \bibnamefont {Botton}},
  \bibinfo {author} {\bibfnamefont {S.~Y.}\ \bibnamefont {Savrasov}}, \bibinfo
  {author} {\bibfnamefont {C.~J.}\ \bibnamefont {Humphreys}},\ and\ \bibinfo
  {author} {\bibfnamefont {A.~P.}\ \bibnamefont {Sutton}},\ }\bibfield  {title}
  {\bibinfo {title} {Electron-energy-loss spectra and the structural stability
  of nickel oxide: An {L}{S}{D}{A}+{U} study},\ }\href@noop {} {\bibfield
  {journal} {\bibinfo  {journal} {Physical Review B}\ }\textbf {\bibinfo
  {volume} {57}},\ \bibinfo {pages} {1505} (\bibinfo {year}
  {1998})}\BibitemShut {NoStop}%
\bibitem [{\citenamefont {Cococcioni}\ and\ \citenamefont
  {de~Gironcoli}(2005)}]{Cococcioni}%
  \BibitemOpen
  \bibfield  {author} {\bibinfo {author} {\bibfnamefont {M.}~\bibnamefont
  {Cococcioni}}\ and\ \bibinfo {author} {\bibfnamefont {S.}~\bibnamefont
  {de~Gironcoli}},\ }\bibfield  {title} {\bibinfo {title} {Linear response
  approach to the calculation of the effective interaction parameters in the
  {L}{D}{A}+{U} method},\ }\href@noop {} {\bibfield  {journal} {\bibinfo
  {journal} {Physical Review B}\ }\textbf {\bibinfo {volume} {71}},\ \bibinfo
  {pages} {035105} (\bibinfo {year} {2005})}\BibitemShut {NoStop}%
\bibitem [{\citenamefont {Giannozzi}\ and\ \citenamefont {et~al.}(2009)}]{QE}%
  \BibitemOpen
  \bibfield  {author} {\bibinfo {author} {\bibfnamefont {P.}~\bibnamefont
  {Giannozzi}}\ and\ \bibinfo {author} {\bibnamefont {et~al.}},\ }\bibfield
  {title} {\bibinfo {title} {Quantum espresso: a modular and open-source
  software project for quantum simulations of materials},\ }\href@noop {}
  {\bibfield  {journal} {\bibinfo  {journal} {Journal of physics: Condensed
  matter}\ }\textbf {\bibinfo {volume} {39}},\ \bibinfo {pages} {395502}
  (\bibinfo {year} {2009})}\BibitemShut {NoStop}%
\bibitem [{\citenamefont {et~al.}(2017)}]{QE2}%
  \BibitemOpen
  \bibfield  {author} {\bibinfo {author} {\bibfnamefont {P.~G.}\ \bibnamefont
  {et~al.}},\ }\bibfield  {title} {\bibinfo {title} {Advanced capabilities for
  materials modelling with quantum {E}{S}{P}{R}{E}{S}{S}{O}},\ }\href@noop {}
  {\bibfield  {journal} {\bibinfo  {journal} {J. Phys.: Condens. Matter}\
  }\textbf {\bibinfo {volume} {29}},\ \bibinfo {pages} {465901} (\bibinfo
  {year} {2017})}\BibitemShut {NoStop}%
\bibitem [{\citenamefont {Corso}(2014)}]{UPF}%
  \BibitemOpen
  \bibfield  {author} {\bibinfo {author} {\bibfnamefont {A.~D.}\ \bibnamefont
  {Corso}},\ }\bibfield  {title} {\bibinfo {title} {Pseudopotentials periodic
  table: From h to pu},\ }\href@noop {} {\bibfield  {journal} {\bibinfo
  {journal} {Computational Materials Science}\ }\textbf {\bibinfo {volume}
  {95}},\ \bibinfo {pages} {337} (\bibinfo {year} {2014})}\BibitemShut
  {NoStop}%
\bibitem [{\citenamefont {Perdew}\ \emph {et~al.}(2008)\citenamefont {Perdew},
  \citenamefont {Ruzsinszky}, \citenamefont {Csonka}, \citenamefont {Vydrov},
  \citenamefont {Scuseria}, \citenamefont {Constantin}, \citenamefont {Zhou},\
  and\ \citenamefont {Burke}}]{PBEsol}%
  \BibitemOpen
  \bibfield  {author} {\bibinfo {author} {\bibfnamefont {J.~P.}\ \bibnamefont
  {Perdew}}, \bibinfo {author} {\bibfnamefont {A.}~\bibnamefont {Ruzsinszky}},
  \bibinfo {author} {\bibfnamefont {G.~I.}\ \bibnamefont {Csonka}}, \bibinfo
  {author} {\bibfnamefont {O.~A.}\ \bibnamefont {Vydrov}}, \bibinfo {author}
  {\bibfnamefont {G.~E.}\ \bibnamefont {Scuseria}}, \bibinfo {author}
  {\bibfnamefont {L.~A.}\ \bibnamefont {Constantin}}, \bibinfo {author}
  {\bibfnamefont {X.}~\bibnamefont {Zhou}},\ and\ \bibinfo {author}
  {\bibfnamefont {K.}~\bibnamefont {Burke}},\ }\bibfield  {title} {\bibinfo
  {title} {Restoring the density-gradient expansion for exchange in solids and
  surfaces},\ }\href@noop {} {\bibfield  {journal} {\bibinfo  {journal}
  {Physical Review Letters}\ }\textbf {\bibinfo {volume} {100}},\ \bibinfo
  {pages} {136406} (\bibinfo {year} {2008})}\BibitemShut {NoStop}%
\bibitem [{\citenamefont {Benjelloun}\ \emph {et~al.}(1984)\citenamefont
  {Benjelloun}, \citenamefont {Bonnet}, \citenamefont {Doumerc}, \citenamefont
  {Launay}, \citenamefont {Onillon},\ and\ \citenamefont
  {Hagenmuller}}]{Benjelloun}%
  \BibitemOpen
  \bibfield  {author} {\bibinfo {author} {\bibfnamefont {D.}~\bibnamefont
  {Benjelloun}}, \bibinfo {author} {\bibfnamefont {J.-P.}\ \bibnamefont
  {Bonnet}}, \bibinfo {author} {\bibfnamefont {J.-P.}\ \bibnamefont {Doumerc}},
  \bibinfo {author} {\bibfnamefont {J.-C.}\ \bibnamefont {Launay}}, \bibinfo
  {author} {\bibfnamefont {M.}~\bibnamefont {Onillon}},\ and\ \bibinfo {author}
  {\bibfnamefont {P.}~\bibnamefont {Hagenmuller}},\ }\bibfield  {title}
  {\bibinfo {title} {Anisotropie des proprietes electriques de l'oxyde de fer
  \ch{Fe2O3}-$\alpha$-},\ }\href@noop {} {\bibfield  {journal} {\bibinfo
  {journal} {Mater. Chem. Phys.}\ }\textbf {\bibinfo {volume} {10}},\ \bibinfo
  {pages} {503} (\bibinfo {year} {1984})}\BibitemShut {NoStop}%
\bibitem [{\citenamefont {Coey}\ and\ \citenamefont {Sawatzky}(1971)}]{Coey}%
  \BibitemOpen
  \bibfield  {author} {\bibinfo {author} {\bibfnamefont {J.~M.~D.}\
  \bibnamefont {Coey}}\ and\ \bibinfo {author} {\bibfnamefont {G.~A.}\
  \bibnamefont {Sawatzky}},\ }\bibfield  {title} {\bibinfo {title} {A study of
  hyperfine interactions in the system \ch{(Fe1-xRhx)2O3} using the mossbauer
  effect (bonding parameters)},\ }\href@noop {} {\bibfield  {journal} {\bibinfo
   {journal} {J. Phys. C}\ }\textbf {\bibinfo {volume} {4}},\ \bibinfo {pages}
  {2386} (\bibinfo {year} {1971})}\BibitemShut {NoStop}%
\bibitem [{\citenamefont {Baroni}\ \emph {et~al.}(73)\citenamefont {Baroni},
  \citenamefont {de~Gironcoli}, \citenamefont {Corso}, ,\ and\ \citenamefont
  {Giannozzi}}]{Baroni}%
  \BibitemOpen
  \bibfield  {author} {\bibinfo {author} {\bibfnamefont {S.}~\bibnamefont
  {Baroni}}, \bibinfo {author} {\bibfnamefont {S.}~\bibnamefont
  {de~Gironcoli}}, \bibinfo {author} {\bibfnamefont {A.~D.}\ \bibnamefont
  {Corso}}, ,\ and\ \bibinfo {author} {\bibfnamefont {P.}~\bibnamefont
  {Giannozzi}},\ }\bibfield  {title} {\bibinfo {title} {Phonons and related
  crystal properties from density-functional perturbation theory},\ }\href@noop
  {} {\bibfield  {journal} {\bibinfo  {journal} {Rev. Mod. Phys.}\ }\textbf
  {\bibinfo {volume} {515}},\ \bibinfo {pages} {2001} (\bibinfo {year}
  {73})}\BibitemShut {NoStop}%
\bibitem [{\citenamefont {Renganathan}\ \emph {et~al.}(2023)\citenamefont
  {Renganathan}, \citenamefont {Sharma}, \citenamefont {Turneaure},\ and\
  \citenamefont {Gupta}}]{renganathan}%
  \BibitemOpen
  \bibfield  {author} {\bibinfo {author} {\bibfnamefont {P.}~\bibnamefont
  {Renganathan}}, \bibinfo {author} {\bibfnamefont {S.~M.}\ \bibnamefont
  {Sharma}}, \bibinfo {author} {\bibfnamefont {S.~J.}\ \bibnamefont
  {Turneaure}},\ and\ \bibinfo {author} {\bibfnamefont {Y.~M.}\ \bibnamefont
  {Gupta}},\ }\bibfield  {title} {\bibinfo {title} {Real-time (nanoseconds)
  determination of liquid phase growth during shock-induced melting},\ }\href
  {https://doi.org/https://doi.org/10.1126/sciadv.ade5745} {\bibfield
  {journal} {\bibinfo  {journal} {Science Advances}\ }\textbf {\bibinfo
  {volume} {9}},\ \bibinfo {pages} {eade5745} (\bibinfo {year}
  {2023})}\BibitemShut {NoStop}%
\bibitem [{\citenamefont {McBride}\ \emph {et~al.}(2019)\citenamefont
  {McBride}, \citenamefont {Krygier}, \citenamefont {Ehnes}, \citenamefont
  {Galtier}, \citenamefont {Harmand}, \citenamefont {Konôpková},
  \citenamefont {Lee}, \citenamefont {Liermann}, \citenamefont {Nagler},
  \citenamefont {Pelka}, \citenamefont {Rödel}, \citenamefont {Schropp},
  \citenamefont {Smith}, \citenamefont {Spindloe}, \citenamefont {Swift},
  \citenamefont {Tavella}, \citenamefont {Toleikis}, \citenamefont
  {Tschentscher}, \citenamefont {Wark},\ and\ \citenamefont
  {Higginbotham}}]{McBride}%
  \BibitemOpen
  \bibfield  {author} {\bibinfo {author} {\bibfnamefont {E.~E.}\ \bibnamefont
  {McBride}}, \bibinfo {author} {\bibfnamefont {A.}~\bibnamefont {Krygier}},
  \bibinfo {author} {\bibfnamefont {A.}~\bibnamefont {Ehnes}}, \bibinfo
  {author} {\bibfnamefont {E.}~\bibnamefont {Galtier}}, \bibinfo {author}
  {\bibfnamefont {M.}~\bibnamefont {Harmand}}, \bibinfo {author} {\bibfnamefont
  {Z.}~\bibnamefont {Konôpková}}, \bibinfo {author} {\bibfnamefont {H.~J.}\
  \bibnamefont {Lee}}, \bibinfo {author} {\bibfnamefont {H.-P.}\ \bibnamefont
  {Liermann}}, \bibinfo {author} {\bibfnamefont {B.}~\bibnamefont {Nagler}},
  \bibinfo {author} {\bibfnamefont {A.}~\bibnamefont {Pelka}}, \bibinfo
  {author} {\bibfnamefont {M.}~\bibnamefont {Rödel}}, \bibinfo {author}
  {\bibfnamefont {A.}~\bibnamefont {Schropp}}, \bibinfo {author} {\bibfnamefont
  {R.~F.}\ \bibnamefont {Smith}}, \bibinfo {author} {\bibfnamefont
  {C.}~\bibnamefont {Spindloe}}, \bibinfo {author} {\bibfnamefont
  {D.}~\bibnamefont {Swift}}, \bibinfo {author} {\bibfnamefont
  {F.}~\bibnamefont {Tavella}}, \bibinfo {author} {\bibfnamefont
  {S.}~\bibnamefont {Toleikis}}, \bibinfo {author} {\bibfnamefont
  {T.}~\bibnamefont {Tschentscher}}, \bibinfo {author} {\bibfnamefont {J.~S.}\
  \bibnamefont {Wark}},\ and\ \bibinfo {author} {\bibfnamefont
  {A.}~\bibnamefont {Higginbotham}},\ }\bibfield  {title} {\bibinfo {title}
  {Phase transition lowering in dynamically compressed silicon},\ }\href
  {https://doi.org/https://doi.org/10.1038/s41567-018-0290-x} {\bibfield
  {journal} {\bibinfo  {journal} {Nature Physics}\ }\textbf {\bibinfo {volume}
  {15}},\ \bibinfo {pages} {89} (\bibinfo {year} {2019})}\BibitemShut {NoStop}%
\bibitem [{\citenamefont {Shi}\ \emph {et~al.}(2020)\citenamefont {Shi},
  \citenamefont {Alderman}, \citenamefont {Tamalonis}, \citenamefont {Weber},
  \citenamefont {You},\ and\ \citenamefont {Benmore}}]{shi}%
  \BibitemOpen
  \bibfield  {author} {\bibinfo {author} {\bibfnamefont {C.}~\bibnamefont
  {Shi}}, \bibinfo {author} {\bibfnamefont {O.~L.~G.}\ \bibnamefont
  {Alderman}}, \bibinfo {author} {\bibfnamefont {A.}~\bibnamefont {Tamalonis}},
  \bibinfo {author} {\bibfnamefont {R.}~\bibnamefont {Weber}}, \bibinfo
  {author} {\bibfnamefont {J.}~\bibnamefont {You}},\ and\ \bibinfo {author}
  {\bibfnamefont {C.~J.}\ \bibnamefont {Benmore}},\ }\bibfield  {title}
  {\bibinfo {title} {Redox-structure dependence of molten iron oxides},\ }\href
  {https://doi.org/https://doi.org/10.1038/s43246-020-00080-4} {\bibfield
  {journal} {\bibinfo  {journal} {Communications Materials}\ }\textbf {\bibinfo
  {volume} {1}},\ \bibinfo {pages} {80} (\bibinfo {year} {2020})}\BibitemShut
  {NoStop}%
\bibitem [{\citenamefont {Morard}\ \emph {et~al.}(2022)\citenamefont {Morard},
  \citenamefont {Antonangeli}, \citenamefont {Bouchet}, \citenamefont
  {Rivoldini}, \citenamefont {Boccato}, \citenamefont {Miozzi}, \citenamefont
  {Boulard}, \citenamefont {Bureau}, \citenamefont {Mezouar}, \citenamefont
  {Prescher}, \citenamefont {Chariton},\ and\ \citenamefont
  {Greenberg}}]{morard2022}%
  \BibitemOpen
  \bibfield  {author} {\bibinfo {author} {\bibfnamefont {G.}~\bibnamefont
  {Morard}}, \bibinfo {author} {\bibfnamefont {D.}~\bibnamefont {Antonangeli}},
  \bibinfo {author} {\bibfnamefont {J.}~\bibnamefont {Bouchet}}, \bibinfo
  {author} {\bibfnamefont {A.}~\bibnamefont {Rivoldini}}, \bibinfo {author}
  {\bibfnamefont {S.}~\bibnamefont {Boccato}}, \bibinfo {author} {\bibfnamefont
  {F.}~\bibnamefont {Miozzi}}, \bibinfo {author} {\bibfnamefont
  {E.}~\bibnamefont {Boulard}}, \bibinfo {author} {\bibfnamefont
  {H.}~\bibnamefont {Bureau}}, \bibinfo {author} {\bibfnamefont
  {M.}~\bibnamefont {Mezouar}}, \bibinfo {author} {\bibfnamefont
  {C.}~\bibnamefont {Prescher}}, \bibinfo {author} {\bibfnamefont
  {S.}~\bibnamefont {Chariton}},\ and\ \bibinfo {author} {\bibfnamefont
  {E.}~\bibnamefont {Greenberg}},\ }\bibfield  {title} {\bibinfo {title}
  {Structural and {E}lectronic {T}ransitions in {L}iquid {F}e{O} {U}nder {H}igh
  {P}ressure},\ }\href {https://doi.org/https://doi.org/10.1029/2022JB025117}
  {\bibfield  {journal} {\bibinfo  {journal} {Journal of Geophysical Research:
  Solid Earth}\ }\textbf {\bibinfo {volume} {127}},\ \bibinfo {pages}
  {e2022JB025117} (\bibinfo {year} {2022})}\BibitemShut {NoStop}%
\bibitem [{\citenamefont {Raikes}\ and\ \citenamefont {Ahrens}(1979)}]{Raikes}%
  \BibitemOpen
  \bibfield  {author} {\bibinfo {author} {\bibfnamefont {S.}~\bibnamefont
  {Raikes}}\ and\ \bibinfo {author} {\bibfnamefont {T.}~\bibnamefont
  {Ahrens}},\ }\bibfield  {title} {\bibinfo {title} {Post-shock temperatures in
  minerals},\ }\href {https://doi.org/10.1111/j.1365-246X.1979.tb04804.x}
  {\bibfield  {journal} {\bibinfo  {journal} {Geophysical Journal
  International}\ }\textbf {\bibinfo {volume} {58}},\ \bibinfo {pages}
  {717–747} (\bibinfo {year} {1979})}\BibitemShut {NoStop}%
\bibitem [{\citenamefont {Sharp}\ and\ \citenamefont {DeCarli}(2006)}]{Sharp}%
  \BibitemOpen
  \bibfield  {author} {\bibinfo {author} {\bibfnamefont {T.}~\bibnamefont
  {Sharp}}\ and\ \bibinfo {author} {\bibfnamefont {P.}~\bibnamefont
  {DeCarli}},\ }\bibfield  {title} {\bibinfo {title} {Shock effects in
  meteorites},\ }\href {https://doi.org/10.2307/j.ctv1v7zdmm.37} {\bibfield
  {journal} {\bibinfo  {journal} {in Meteorites and the Early Solar System II
  Space Science Series}\ ,\ \bibinfo {pages} {653}} (\bibinfo {year}
  {2006})}\BibitemShut {NoStop}%
\bibitem [{\citenamefont {Kalita}\ \emph {et~al.}(2023)\citenamefont {Kalita},
  \citenamefont {Cochrane}, \citenamefont {Knudson}, \citenamefont {Ao},
  \citenamefont {Blada}, \citenamefont {Jackson}, \citenamefont {Gluth},
  \citenamefont {Hanshaw}, \citenamefont {Scoglietti},\ and\ \citenamefont
  {Crockett}}]{kalita}%
  \BibitemOpen
  \bibfield  {author} {\bibinfo {author} {\bibfnamefont {P.}~\bibnamefont
  {Kalita}}, \bibinfo {author} {\bibfnamefont {K.~R.}\ \bibnamefont
  {Cochrane}}, \bibinfo {author} {\bibfnamefont {M.~D.}\ \bibnamefont
  {Knudson}}, \bibinfo {author} {\bibfnamefont {T.}~\bibnamefont {Ao}},
  \bibinfo {author} {\bibfnamefont {C.}~\bibnamefont {Blada}}, \bibinfo
  {author} {\bibfnamefont {J.}~\bibnamefont {Jackson}}, \bibinfo {author}
  {\bibfnamefont {J.}~\bibnamefont {Gluth}}, \bibinfo {author} {\bibfnamefont
  {H.}~\bibnamefont {Hanshaw}}, \bibinfo {author} {\bibfnamefont
  {E.}~\bibnamefont {Scoglietti}},\ and\ \bibinfo {author} {\bibfnamefont
  {S.~D.}\ \bibnamefont {Crockett}},\ }\bibfield  {title} {\bibinfo {title}
  {{Ti-6Al-4V to over 1.2 TPa: Shock Hugoniot experiments, ab initio
  calculations, and a broad-range multiphase equation of state}},\ }\href
  {https://doi.org/10.1103/PhysRevLett.89.205504} {\bibfield  {journal}
  {\bibinfo  {journal} {Physical Review B.}\ }\textbf {\bibinfo {volume}
  {107}},\ \bibinfo {pages} {094101} (\bibinfo {year} {2023})}\BibitemShut
  {NoStop}%
\bibitem [{\citenamefont {Biswas}(1981)}]{biswas}%
  \BibitemOpen
  \bibfield  {author} {\bibinfo {author} {\bibfnamefont {A.~K.}\ \bibnamefont
  {Biswas}},\ }\bibfield  {title} {\bibinfo {title} {Principles of {B}last
  {F}urnace {I}ronmaking : {T}heory and {P}ractice}\ }(\bibinfo  {publisher}
  {Cootha Publishing House Brisbane, Australia},\ \bibinfo {year}
  {1981})\BibitemShut {NoStop}%
\bibitem [{\citenamefont {Deng}\ \emph {et~al.}(2019)\citenamefont {Deng},
  \citenamefont {Karki}, \citenamefont {Ghosh},\ and\ \citenamefont
  {Lee}}]{deng}%
  \BibitemOpen
  \bibfield  {author} {\bibinfo {author} {\bibfnamefont {J.}~\bibnamefont
  {Deng}}, \bibinfo {author} {\bibfnamefont {B.~B.}\ \bibnamefont {Karki}},
  \bibinfo {author} {\bibfnamefont {D.~B.}\ \bibnamefont {Ghosh}},\ and\
  \bibinfo {author} {\bibfnamefont {K.~K.~M.}\ \bibnamefont {Lee}},\ }\bibfield
   {title} {\bibinfo {title} {First-{P}rinciples {S}tudy of \ch{FeO2Hx} {S}olid
  and {M}elt {S}ystem at {H}igh {P}ressures: {I}mplications for
  {U}ltralow-{V}elocity {Z}ones},\ }\href
  {https://doi.org/https://doi.org/10.1029/2019JB017376} {\bibfield  {journal}
  {\bibinfo  {journal} {Journal of Geophysical Research: Solid Earth}\ }\textbf
  {\bibinfo {volume} {124}},\ \bibinfo {pages} {4566} (\bibinfo {year}
  {2019})}\BibitemShut {NoStop}%
\bibitem [{\citenamefont {Jeanloz}\ and\ \citenamefont
  {Ahrens}(1980)}]{jeanloz}%
  \BibitemOpen
  \bibfield  {author} {\bibinfo {author} {\bibfnamefont {R.}~\bibnamefont
  {Jeanloz}}\ and\ \bibinfo {author} {\bibfnamefont {T.~J.}\ \bibnamefont
  {Ahrens}},\ }\bibfield  {title} {\bibinfo {title} {Equations of state of
  {F}e{O} and {C}a{O}},\ }\href
  {https://doi.org/https://doi.org/10.1111/j.1365-246X.1980.tb02588.x}
  {\bibfield  {journal} {\bibinfo  {journal} {Geophysical Journal
  International}\ }\textbf {\bibinfo {volume} {62}},\ \bibinfo {pages}
  {505–528} (\bibinfo {year} {1980})}\BibitemShut {NoStop}%
\bibitem [{\citenamefont {Boehler}(1993)}]{boehler1993}%
  \BibitemOpen
  \bibfield  {author} {\bibinfo {author} {\bibfnamefont {R.}~\bibnamefont
  {Boehler}},\ }\bibfield  {title} {\bibinfo {title} {Temperatures in the
  {E}arth's core from melting-point measurements of iron at high static
  pressures},\ }\href {https://doi.org/https://doi.org/10.1038/363534a0}
  {\bibfield  {journal} {\bibinfo  {journal} {Nature}\ }\textbf {\bibinfo
  {volume} {363}},\ \bibinfo {pages} {534–536} (\bibinfo {year}
  {1993})}\BibitemShut {NoStop}%
\bibitem [{\citenamefont {Anzellini}\ \emph {et~al.}(2013)\citenamefont
  {Anzellini}, \citenamefont {Dewaele}, \citenamefont {Mezouar}, \citenamefont
  {Loubeyre},\ and\ \citenamefont {Morard}}]{anzellini}%
  \BibitemOpen
  \bibfield  {author} {\bibinfo {author} {\bibfnamefont {S.}~\bibnamefont
  {Anzellini}}, \bibinfo {author} {\bibfnamefont {A.}~\bibnamefont {Dewaele}},
  \bibinfo {author} {\bibfnamefont {M.}~\bibnamefont {Mezouar}}, \bibinfo
  {author} {\bibfnamefont {P.}~\bibnamefont {Loubeyre}},\ and\ \bibinfo
  {author} {\bibfnamefont {G.}~\bibnamefont {Morard}},\ }\bibfield  {title}
  {\bibinfo {title} {Melting of {I}ron at {E}arth’s {I}nner {C}ore {B}oundary
  {B}ased on {F}ast {X}-ray {D}iffraction},\ }\href
  {https://doi.org/https://doi.org/10.1126/science.1233514} {\bibfield
  {journal} {\bibinfo  {journal} {Science}\ }\textbf {\bibinfo {volume}
  {340}},\ \bibinfo {pages} {464} (\bibinfo {year} {2013})}\BibitemShut
  {NoStop}%
\bibitem [{\citenamefont {Dobrosavljevic}\ \emph {et~al.}(2023)\citenamefont
  {Dobrosavljevic}, \citenamefont {Zhang}, \citenamefont {Sturhahn},
  \citenamefont {Chariton}, \citenamefont {Prakapenka}, \citenamefont {Zhao},
  \citenamefont {Toellner}, \citenamefont {Pardo},\ and\ \citenamefont
  {Jackson}}]{dobrosavljevic}%
  \BibitemOpen
  \bibfield  {author} {\bibinfo {author} {\bibfnamefont {V.~V.}\ \bibnamefont
  {Dobrosavljevic}}, \bibinfo {author} {\bibfnamefont {D.}~\bibnamefont
  {Zhang}}, \bibinfo {author} {\bibfnamefont {W.}~\bibnamefont {Sturhahn}},
  \bibinfo {author} {\bibfnamefont {S.}~\bibnamefont {Chariton}}, \bibinfo
  {author} {\bibfnamefont {V.~B.}\ \bibnamefont {Prakapenka}}, \bibinfo
  {author} {\bibfnamefont {J.}~\bibnamefont {Zhao}}, \bibinfo {author}
  {\bibfnamefont {T.~S.}\ \bibnamefont {Toellner}}, \bibinfo {author}
  {\bibfnamefont {O.~S.}\ \bibnamefont {Pardo}},\ and\ \bibinfo {author}
  {\bibfnamefont {J.~M.}\ \bibnamefont {Jackson}},\ }\bibfield  {title}
  {\bibinfo {title} {Melting and defect transitions in {F}e{O} up to pressures
  of {E}arth’s core-mantle boundary},\ }\href
  {https://doi.org/https://doi.org/10.1038/s41467-023-43154-w} {\bibfield
  {journal} {\bibinfo  {journal} {Nature Communications}\ }\textbf {\bibinfo
  {volume} {14}},\ \bibinfo {pages} {7336} (\bibinfo {year}
  {2023})}\BibitemShut {NoStop}%
\bibitem [{\citenamefont {Amouretti}\ \emph
  {et~al.}(2024{\natexlab{b}})\citenamefont {Amouretti}, \citenamefont
  {Harmand}, \citenamefont {Albertazzi}, \citenamefont {Boury}, \citenamefont
  {Benuzzi-Mounaix}, \citenamefont {Chin}, \citenamefont {Guyot}, \citenamefont
  {Koenig}, \citenamefont {Vinci},\ and\ \citenamefont {Fiquet}}]{LULI}%
  \BibitemOpen
  \bibfield  {author} {\bibinfo {author} {\bibfnamefont {A.}~\bibnamefont
  {Amouretti}}, \bibinfo {author} {\bibfnamefont {M.}~\bibnamefont {Harmand}},
  \bibinfo {author} {\bibfnamefont {B.}~\bibnamefont {Albertazzi}}, \bibinfo
  {author} {\bibfnamefont {A.}~\bibnamefont {Boury}}, \bibinfo {author}
  {\bibfnamefont {A.}~\bibnamefont {Benuzzi-Mounaix}}, \bibinfo {author}
  {\bibfnamefont {D.~A.}\ \bibnamefont {Chin}}, \bibinfo {author}
  {\bibfnamefont {F.}~\bibnamefont {Guyot}}, \bibinfo {author} {\bibfnamefont
  {M.}~\bibnamefont {Koenig}}, \bibinfo {author} {\bibfnamefont
  {T.}~\bibnamefont {Vinci}},\ and\ \bibinfo {author} {\bibfnamefont
  {G.}~\bibnamefont {Fiquet}},\ }\bibfield  {title} {\bibinfo {title}
  {Laser-driven shock compression and equation of state of \ch{Fe2O3} up to 700
  {G}{P}a}\ }\href {https://doi.org/10.48550/arXiv.2405.08350}
  {10.48550/arXiv.2405.08350} (\bibinfo {year}
  {2024}{\natexlab{b}})\BibitemShut {NoStop}%
\bibitem [{\citenamefont {Idrissi}\ \emph {et~al.}(2022)\citenamefont
  {Idrissi}, \citenamefont {Carrez},\ and\ \citenamefont {Cordier}}]{idrissi}%
  \BibitemOpen
  \bibfield  {author} {\bibinfo {author} {\bibfnamefont {H.}~\bibnamefont
  {Idrissi}}, \bibinfo {author} {\bibfnamefont {P.}~\bibnamefont {Carrez}},\
  and\ \bibinfo {author} {\bibfnamefont {P.}~\bibnamefont {Cordier}},\
  }\bibfield  {title} {\bibinfo {title} {On amorphization as a deformation
  mechanism under high stresses},\ }\href
  {https://doi.org/https://doi.org/10.1016/j.cossms.2021.100976} {\bibfield
  {journal} {\bibinfo  {journal} {Current Opinion in Solid State and Materials
  Science}\ }\textbf {\bibinfo {volume} {26}},\ \bibinfo {pages} {100976}
  (\bibinfo {year} {2022})}\BibitemShut {NoStop}%
\bibitem [{\citenamefont {Li}\ \emph {et~al.}(2022)\citenamefont {Li},
  \citenamefont {Li}, \citenamefont {Zhao},\ and\ \citenamefont {Meyers}}]{li}%
  \BibitemOpen
  \bibfield  {author} {\bibinfo {author} {\bibfnamefont {B.}~\bibnamefont
  {Li}}, \bibinfo {author} {\bibfnamefont {A.}~\bibnamefont {Li}}, \bibinfo
  {author} {\bibfnamefont {S.}~\bibnamefont {Zhao}},\ and\ \bibinfo {author}
  {\bibfnamefont {M.}~\bibnamefont {Meyers}},\ }\bibfield  {title} {\bibinfo
  {title} {{Amorphization by mechanical deformation}},\ }\href
  {https://doi.org/10.1016/j.mser.2022.100673} {\bibfield  {journal} {\bibinfo
  {journal} {Materials Science and Engineering: R: Reports}\ }\textbf {\bibinfo
  {volume} {108}},\ \bibinfo {pages} {100673} (\bibinfo {year}
  {2022})}\BibitemShut {NoStop}%
\bibitem [{\citenamefont {Tracy}\ \emph {et~al.}(2018)\citenamefont {Tracy},
  \citenamefont {Turneaure},\ and\ \citenamefont {Duffy}}]{fused_silica}%
  \BibitemOpen
  \bibfield  {author} {\bibinfo {author} {\bibfnamefont {S.~J.}\ \bibnamefont
  {Tracy}}, \bibinfo {author} {\bibfnamefont {S.~J.}\ \bibnamefont
  {Turneaure}},\ and\ \bibinfo {author} {\bibfnamefont {T.~S.}\ \bibnamefont
  {Duffy}},\ }\bibfield  {title} {\bibinfo {title} {{In situ X-Ray Diffraction
  of Shock-Compressed Fused Silica}},\ }\href
  {https://doi.org/10.1103/PhysRevLett.120.135702} {\bibfield  {journal}
  {\bibinfo  {journal} {Physical Review Letters}\ }\textbf {\bibinfo {volume}
  {120}},\ \bibinfo {pages} {135702} (\bibinfo {year} {2018})}\BibitemShut
  {NoStop}%
\bibitem [{\citenamefont {Gleason}\ \emph {et~al.}(2022)\citenamefont
  {Gleason}, \citenamefont {Park}, \citenamefont {Rittman}, \citenamefont
  {Ravasio}, \citenamefont {Langenhorst}, \citenamefont {Bolis}, \citenamefont
  {Granados}, \citenamefont {Hok}, \citenamefont {Kroll}, \citenamefont
  {Sikorski}, \citenamefont {Weng}, \citenamefont {Lee}, \citenamefont
  {Nagler}, \citenamefont {Sisson}, \citenamefont {Xing}, \citenamefont {Zhu},
  \citenamefont {Giuli}, \citenamefont {Mao}, \citenamefont {Glenzer},
  \citenamefont {Sokaras},\ and\ \citenamefont {Alonso‐Mori}}]{gleason}%
  \BibitemOpen
  \bibfield  {author} {\bibinfo {author} {\bibfnamefont {A.~E.}\ \bibnamefont
  {Gleason}}, \bibinfo {author} {\bibfnamefont {S.}~\bibnamefont {Park}},
  \bibinfo {author} {\bibfnamefont {D.~R.}\ \bibnamefont {Rittman}}, \bibinfo
  {author} {\bibfnamefont {A.}~\bibnamefont {Ravasio}}, \bibinfo {author}
  {\bibfnamefont {F.}~\bibnamefont {Langenhorst}}, \bibinfo {author}
  {\bibfnamefont {R.~M.}\ \bibnamefont {Bolis}}, \bibinfo {author}
  {\bibfnamefont {E.}~\bibnamefont {Granados}}, \bibinfo {author}
  {\bibfnamefont {S.}~\bibnamefont {Hok}}, \bibinfo {author} {\bibfnamefont
  {T.}~\bibnamefont {Kroll}}, \bibinfo {author} {\bibfnamefont
  {M.}~\bibnamefont {Sikorski}}, \bibinfo {author} {\bibfnamefont {T.-C.}\
  \bibnamefont {Weng}}, \bibinfo {author} {\bibfnamefont {H.~J.}\ \bibnamefont
  {Lee}}, \bibinfo {author} {\bibfnamefont {B.}~\bibnamefont {Nagler}},
  \bibinfo {author} {\bibfnamefont {T.}~\bibnamefont {Sisson}}, \bibinfo
  {author} {\bibfnamefont {Z.}~\bibnamefont {Xing}}, \bibinfo {author}
  {\bibfnamefont {D.}~\bibnamefont {Zhu}}, \bibinfo {author} {\bibfnamefont
  {G.}~\bibnamefont {Giuli}}, \bibinfo {author} {\bibfnamefont {W.~L.}\
  \bibnamefont {Mao}}, \bibinfo {author} {\bibfnamefont {S.}~\bibnamefont
  {Glenzer}}, \bibinfo {author} {\bibfnamefont {D.}~\bibnamefont {Sokaras}},\
  and\ \bibinfo {author} {\bibfnamefont {R.}~\bibnamefont {Alonso‐Mori}},\
  }\bibfield  {title} {\bibinfo {title} {{Ultrafast structural response of
  shock-compressed plagioclase}},\ }\href {https://doi.org/10.1111/maps.13785}
  {\bibfield  {journal} {\bibinfo  {journal} {Meteoritics and Planetary
  Science}\ }\textbf {\bibinfo {volume} {57}},\ \bibinfo {pages} {635–643}
  (\bibinfo {year} {2022})}\BibitemShut {NoStop}%
\bibitem [{\citenamefont {Hernandez}\ \emph {et~al.}(2020)\citenamefont
  {Hernandez}, \citenamefont {Morard}, \citenamefont {Guarguaglini},
  \citenamefont {Alonso‐Mori}, \citenamefont {Benuzzi‐Mounaix},
  \citenamefont {Bolis}, \citenamefont {Fiquet}, \citenamefont {Galtier},
  \citenamefont {Gleason}, \citenamefont {Glenzer}, \citenamefont {Guyot},
  \citenamefont {Ko}, \citenamefont {Lee}, \citenamefont {Mao}, \citenamefont
  {Nagler}, \citenamefont {Ozaki}, \citenamefont {Schuster}, \citenamefont
  {Shim}, \citenamefont {Vinci},\ and\ \citenamefont {Ravasio1}}]{hernandez}%
  \BibitemOpen
  \bibfield  {author} {\bibinfo {author} {\bibfnamefont {J.}~\bibnamefont
  {Hernandez}}, \bibinfo {author} {\bibfnamefont {G.}~\bibnamefont {Morard}},
  \bibinfo {author} {\bibfnamefont {M.}~\bibnamefont {Guarguaglini}}, \bibinfo
  {author} {\bibfnamefont {R.}~\bibnamefont {Alonso‐Mori}}, \bibinfo {author}
  {\bibfnamefont {A.}~\bibnamefont {Benuzzi‐Mounaix}}, \bibinfo {author}
  {\bibfnamefont {R.}~\bibnamefont {Bolis}}, \bibinfo {author} {\bibfnamefont
  {G.}~\bibnamefont {Fiquet}}, \bibinfo {author} {\bibfnamefont
  {E.}~\bibnamefont {Galtier}}, \bibinfo {author} {\bibfnamefont {A.~E.}\
  \bibnamefont {Gleason}}, \bibinfo {author} {\bibfnamefont {S.}~\bibnamefont
  {Glenzer}}, \bibinfo {author} {\bibfnamefont {F.}~\bibnamefont {Guyot}},
  \bibinfo {author} {\bibfnamefont {B.}~\bibnamefont {Ko}}, \bibinfo {author}
  {\bibfnamefont {H.~J.}\ \bibnamefont {Lee}}, \bibinfo {author} {\bibfnamefont
  {W.~L.}\ \bibnamefont {Mao}}, \bibinfo {author} {\bibfnamefont
  {B.}~\bibnamefont {Nagler}}, \bibinfo {author} {\bibfnamefont
  {N.}~\bibnamefont {Ozaki}}, \bibinfo {author} {\bibfnamefont {A.~K.}\
  \bibnamefont {Schuster}}, \bibinfo {author} {\bibfnamefont {S.~H.}\
  \bibnamefont {Shim}}, \bibinfo {author} {\bibfnamefont {T.}~\bibnamefont
  {Vinci}},\ and\ \bibinfo {author} {\bibfnamefont {A.}~\bibnamefont
  {Ravasio1}},\ }\bibfield  {title} {\bibinfo {title} {Direct observation of
  shock‐induced disordering of enstatite below the melting temperature},\
  }\href {https://doi.org/https://doi.org/10.1029/2020GL088887} {\bibfield
  {journal} {\bibinfo  {journal} {Geophysical Research Letter}\ }\textbf
  {\bibinfo {volume} {47}},\ \bibinfo {pages} {e2020GL088887} (\bibinfo {year}
  {2020})}\BibitemShut {NoStop}%
\bibitem [{\citenamefont {Jeanloz}\ \emph {et~al.}(1977)\citenamefont
  {Jeanloz}, \citenamefont {Ahrens}, \citenamefont {Lally}, \citenamefont
  {Nord}, \citenamefont {Christie},\ and\ \citenamefont
  {Heuer}}]{jeanloz_1977}%
  \BibitemOpen
  \bibfield  {author} {\bibinfo {author} {\bibfnamefont {R.}~\bibnamefont
  {Jeanloz}}, \bibinfo {author} {\bibfnamefont {T.~J.}\ \bibnamefont {Ahrens}},
  \bibinfo {author} {\bibfnamefont {J.}~\bibnamefont {Lally}}, \bibinfo
  {author} {\bibfnamefont {G.~L.}\ \bibnamefont {Nord}}, \bibinfo {author}
  {\bibfnamefont {J.~J.~M.}\ \bibnamefont {Christie}},\ and\ \bibinfo {author}
  {\bibfnamefont {A.~H.}\ \bibnamefont {Heuer}},\ }\bibfield  {title} {\bibinfo
  {title} {{Shock-Produced Olivine Glass: First Observation}},\ }\href
  {https://doi.org/10.1126/science.197.4302.457} {\bibfield  {journal}
  {\bibinfo  {journal} {Physical Review B.}\ }\textbf {\bibinfo {volume}
  {197}},\ \bibinfo {pages} {457} (\bibinfo {year} {1977})}\BibitemShut
  {NoStop}%
\bibitem [{\citenamefont {Kim}\ \emph {et~al.}(2020)\citenamefont {Kim},
  \citenamefont {Tracy}, \citenamefont {Smith}, \citenamefont {Gleason},
  \citenamefont {Bolme}, \citenamefont {Prakapenka}, \citenamefont {Appel},
  \citenamefont {Speziale}, \citenamefont {Wicks}, \citenamefont {Berryman},
  \citenamefont {Han}, \citenamefont {Schoelmerich}, \citenamefont {Lee},
  \citenamefont {Nagler}, \citenamefont {Cunningham}, \citenamefont {Akin},
  \citenamefont {Asimow}, \citenamefont {Eggert},\ and\ \citenamefont
  {Duffy}}]{kim2020}%
  \BibitemOpen
  \bibfield  {author} {\bibinfo {author} {\bibfnamefont {D.}~\bibnamefont
  {Kim}}, \bibinfo {author} {\bibfnamefont {S.~J.}\ \bibnamefont {Tracy}},
  \bibinfo {author} {\bibfnamefont {R.~F.}\ \bibnamefont {Smith}}, \bibinfo
  {author} {\bibfnamefont {A.~E.}\ \bibnamefont {Gleason}}, \bibinfo {author}
  {\bibfnamefont {C.~A.}\ \bibnamefont {Bolme}}, \bibinfo {author}
  {\bibfnamefont {V.~B.}\ \bibnamefont {Prakapenka}}, \bibinfo {author}
  {\bibfnamefont {K.}~\bibnamefont {Appel}}, \bibinfo {author} {\bibfnamefont
  {S.}~\bibnamefont {Speziale}}, \bibinfo {author} {\bibfnamefont {J.~K.}\
  \bibnamefont {Wicks}}, \bibinfo {author} {\bibfnamefont {E.~J.}\ \bibnamefont
  {Berryman}}, \bibinfo {author} {\bibfnamefont {S.~K.}\ \bibnamefont {Han}},
  \bibinfo {author} {\bibfnamefont {M.~O.}\ \bibnamefont {Schoelmerich}},
  \bibinfo {author} {\bibfnamefont {H.~J.}\ \bibnamefont {Lee}}, \bibinfo
  {author} {\bibfnamefont {B.}~\bibnamefont {Nagler}}, \bibinfo {author}
  {\bibfnamefont {E.~F.}\ \bibnamefont {Cunningham}}, \bibinfo {author}
  {\bibfnamefont {M.~C.}\ \bibnamefont {Akin}}, \bibinfo {author}
  {\bibfnamefont {P.~D.}\ \bibnamefont {Asimow}}, \bibinfo {author}
  {\bibfnamefont {J.~H.}\ \bibnamefont {Eggert}},\ and\ \bibinfo {author}
  {\bibfnamefont {T.~S.}\ \bibnamefont {Duffy}},\ }\bibfield  {title} {\bibinfo
  {title} {{Femtosecond X-Ray Diffraction of Laser-Shocked Forsterite
  (\ch{Mg2SiO4}) to 122 GPa}},\ }\href {https://doi.org/10.1029/2020JB020337}
  {\bibfield  {journal} {\bibinfo  {journal} {Journal of Geophysical Research:
  Solid Earth}\ }\textbf {\bibinfo {volume} {126}},\ \bibinfo {pages}
  {e2020JB020337} (\bibinfo {year} {2020})}\BibitemShut {NoStop}%
\bibitem [{\citenamefont {Shim}\ \emph {et~al.}(2023)\citenamefont {Shim},
  \citenamefont {Ko}, \citenamefont {Sokaras}, \citenamefont {Nagler},
  \citenamefont {Lee}, \citenamefont {Galtier}, \citenamefont {Glenzer},
  \citenamefont {Granados}, \citenamefont {Vinci}, \citenamefont {Fiquet},
  \citenamefont {Dolinschi}, \citenamefont {Tappan}, \citenamefont {Kulka},
  \citenamefont {Mao}, \citenamefont {Morard}, \citenamefont {Ravasio},
  \citenamefont {Gleason},\ and\ \citenamefont {Alonso-Mori}}]{shim2023}%
  \BibitemOpen
  \bibfield  {author} {\bibinfo {author} {\bibfnamefont {S.-H.}\ \bibnamefont
  {Shim}}, \bibinfo {author} {\bibfnamefont {B.}~\bibnamefont {Ko}}, \bibinfo
  {author} {\bibfnamefont {D.}~\bibnamefont {Sokaras}}, \bibinfo {author}
  {\bibfnamefont {B.}~\bibnamefont {Nagler}}, \bibinfo {author} {\bibfnamefont
  {H.~J.}\ \bibnamefont {Lee}}, \bibinfo {author} {\bibfnamefont
  {E.}~\bibnamefont {Galtier}}, \bibinfo {author} {\bibfnamefont
  {S.}~\bibnamefont {Glenzer}}, \bibinfo {author} {\bibfnamefont
  {E.}~\bibnamefont {Granados}}, \bibinfo {author} {\bibfnamefont
  {T.}~\bibnamefont {Vinci}}, \bibinfo {author} {\bibfnamefont
  {G.}~\bibnamefont {Fiquet}}, \bibinfo {author} {\bibfnamefont
  {J.}~\bibnamefont {Dolinschi}}, \bibinfo {author} {\bibfnamefont
  {J.}~\bibnamefont {Tappan}}, \bibinfo {author} {\bibfnamefont
  {B.}~\bibnamefont {Kulka}}, \bibinfo {author} {\bibfnamefont {W.~L.}\
  \bibnamefont {Mao}}, \bibinfo {author} {\bibfnamefont {G.}~\bibnamefont
  {Morard}}, \bibinfo {author} {\bibfnamefont {A.}~\bibnamefont {Ravasio}},
  \bibinfo {author} {\bibfnamefont {A.}~\bibnamefont {Gleason}},\ and\ \bibinfo
  {author} {\bibfnamefont {R.}~\bibnamefont {Alonso-Mori}},\ }\bibfield
  {title} {\bibinfo {title} {Ultrafast x-ray detection of low-spin iron in
  molten silicate under deep planetary interior conditions},\ }\href
  {https://doi.org/10.1126/sciadv.adi61} {\bibfield  {journal} {\bibinfo
  {journal} {Science Advances}\ }\textbf {\bibinfo {volume} {9}},\ \bibinfo
  {pages} {eadi6153} (\bibinfo {year} {2023})}\BibitemShut {NoStop}%
\bibitem [{\citenamefont {Zhao}\ \emph {et~al.}(2021)\citenamefont {Zhao},
  \citenamefont {Li}, \citenamefont {Remington}, \citenamefont {Wehrenberg},
  \citenamefont {Park}, \citenamefont {Hahn},\ and\ \citenamefont
  {Meyers}}]{zhao}%
  \BibitemOpen
  \bibfield  {author} {\bibinfo {author} {\bibfnamefont {S.}~\bibnamefont
  {Zhao}}, \bibinfo {author} {\bibfnamefont {B.}~\bibnamefont {Li}}, \bibinfo
  {author} {\bibfnamefont {B.}~\bibnamefont {Remington}}, \bibinfo {author}
  {\bibfnamefont {C.}~\bibnamefont {Wehrenberg}}, \bibinfo {author}
  {\bibfnamefont {H.}~\bibnamefont {Park}}, \bibinfo {author} {\bibfnamefont
  {E.}~\bibnamefont {Hahn}},\ and\ \bibinfo {author} {\bibfnamefont
  {M.}~\bibnamefont {Meyers}},\ }\bibfield  {title} {\bibinfo {title}
  {{Directional amorphization of covalently-bonded solids: A generalized
  deformation mechanism in extreme loading}},\ }\href
  {https://doi.org/10.1016/j.mattod.2021.04.017} {\bibfield  {journal}
  {\bibinfo  {journal} {Materials Today}\ }\textbf {\bibinfo {volume} {49}},\
  \bibinfo {pages} {59} (\bibinfo {year} {2021})}\BibitemShut {NoStop}%
\bibitem [{\citenamefont {Liebermann}\ and\ \citenamefont
  {Schreiber}(1968)}]{liberman}%
  \BibitemOpen
  \bibfield  {author} {\bibinfo {author} {\bibfnamefont {R.~C.}\ \bibnamefont
  {Liebermann}}\ and\ \bibinfo {author} {\bibfnamefont {E.}~\bibnamefont
  {Schreiber}},\ }\bibfield  {title} {\bibinfo {title} {Elastic constants of
  polycrystalline hematite as a function of pressure to 3 kilobars},\ }\href
  {https://doi.org/https://doi.org/10.1029/JB073i020p06585} {\bibfield
  {journal} {\bibinfo  {journal} {Journal of Geophysical Research}\ }\textbf
  {\bibinfo {volume} {73}},\ \bibinfo {pages} {6585} (\bibinfo {year}
  {1968})}\BibitemShut {NoStop}%
\bibitem [{\citenamefont {Schouwink}\ \emph {et~al.}(2011)\citenamefont
  {Schouwink}, \citenamefont {Dubrovinsky}, \citenamefont {Glazyrin},
  \citenamefont {Merlini}, \citenamefont {Hanfland}, \citenamefont
  {Pippinger},\ and\ \citenamefont {Miletich}}]{schouwink}%
  \BibitemOpen
  \bibfield  {author} {\bibinfo {author} {\bibfnamefont {P.}~\bibnamefont
  {Schouwink}}, \bibinfo {author} {\bibfnamefont {L.}~\bibnamefont
  {Dubrovinsky}}, \bibinfo {author} {\bibfnamefont {K.}~\bibnamefont
  {Glazyrin}}, \bibinfo {author} {\bibfnamefont {M.}~\bibnamefont {Merlini}},
  \bibinfo {author} {\bibfnamefont {M.}~\bibnamefont {Hanfland}}, \bibinfo
  {author} {\bibfnamefont {T.}~\bibnamefont {Pippinger}},\ and\ \bibinfo
  {author} {\bibfnamefont {R.}~\bibnamefont {Miletich}},\ }\bibfield  {title}
  {\bibinfo {title} {High-pressure structural behavior of $\alpha$-\ch{Fe2O3}
  studied by single-crystal x-ray diffraction and synchrotron radiation up to
  25 {G}{P}a},\ }\href {https://doi.org/https://doi.org/10.2138/am.2011.3775}
  {\bibfield  {journal} {\bibinfo  {journal} {American Mineralogist}\ }\textbf
  {\bibinfo {volume} {96}},\ \bibinfo {pages} {1781–} (\bibinfo {year}
  {2011})}\BibitemShut {NoStop}%
\bibitem [{\citenamefont {Saeki}\ \emph {et~al.}(2011)\citenamefont {Saeki},
  \citenamefont {Ohno}, \citenamefont {Seto}, \citenamefont {Sakai},
  \citenamefont {Sugiyama}, \citenamefont {Sato}, \citenamefont {Yamauchi},
  \citenamefont {Kurokawa}, \citenamefont {Takeda},\ and\ \citenamefont
  {Onishi}}]{saeki}%
  \BibitemOpen
  \bibfield  {author} {\bibinfo {author} {\bibfnamefont {I.}~\bibnamefont
  {Saeki}}, \bibinfo {author} {\bibfnamefont {T.}~\bibnamefont {Ohno}},
  \bibinfo {author} {\bibfnamefont {D.}~\bibnamefont {Seto}}, \bibinfo {author}
  {\bibfnamefont {O.}~\bibnamefont {Sakai}}, \bibinfo {author} {\bibfnamefont
  {Y.}~\bibnamefont {Sugiyama}}, \bibinfo {author} {\bibfnamefont
  {T.}~\bibnamefont {Sato}}, \bibinfo {author} {\bibfnamefont {A.}~\bibnamefont
  {Yamauchi}}, \bibinfo {author} {\bibfnamefont {K.}~\bibnamefont {Kurokawa}},
  \bibinfo {author} {\bibfnamefont {M.}~\bibnamefont {Takeda}},\ and\ \bibinfo
  {author} {\bibfnamefont {T.}~\bibnamefont {Onishi}},\ }\bibfield  {title}
  {\bibinfo {title} {Measurement of young’s modulus of oxides at high
  temperature related to the oxidation study},\ }\href
  {https://doi.org/https://doi.org/10.3184/096034011X13182685579795} {\bibfield
   {journal} {\bibinfo  {journal} {Materials at High Temperatures}\ }\textbf
  {\bibinfo {volume} {28}},\ \bibinfo {pages} {264} (\bibinfo {year}
  {2011})}\BibitemShut {NoStop}%
\bibitem [{\citenamefont {Greer}\ \emph {et~al.}(1975)\citenamefont {Greer},
  \citenamefont {Newham},\ and\ \citenamefont {Pitman}}]{greer}%
  \BibitemOpen
  \bibfield  {author} {\bibinfo {author} {\bibfnamefont {A.}~\bibnamefont
  {Greer}}, \bibinfo {author} {\bibfnamefont {R.}~\bibnamefont {Newham}},\ and\
  \bibinfo {author} {\bibfnamefont {G.}~\bibnamefont {Pitman}},\ }\bibfield
  {title} {\bibinfo {title} {The elastic moduli of single crystals of hematite
  and ilmenite},\ }\href@noop {} {\bibfield  {journal} {\bibinfo  {journal}
  {Journal of Applied Physics}\ }\textbf {\bibinfo {volume} {46}} (\bibinfo
  {year} {1975})}\BibitemShut {NoStop}%
\bibitem [{\citenamefont {Meade}\ \emph {et~al.}(1992)\citenamefont {Meade},
  \citenamefont {Hemley},\ and\ \citenamefont {Mao}}]{meade}%
  \BibitemOpen
  \bibfield  {author} {\bibinfo {author} {\bibfnamefont {C.}~\bibnamefont
  {Meade}}, \bibinfo {author} {\bibfnamefont {R.~J.}\ \bibnamefont {Hemley}},\
  and\ \bibinfo {author} {\bibfnamefont {H.~K.}\ \bibnamefont {Mao}},\
  }\bibfield  {title} {\bibinfo {title} {High-pressure x-ray diffraction of
  \ch{SiO2} glass},\ }\href
  {https://doi.org/https://doi.org/10.1103/PhysRevLett.69.1387} {\bibfield
  {journal} {\bibinfo  {journal} {Physical Review Letters}\ }\textbf {\bibinfo
  {volume} {69}},\ \bibinfo {pages} {1387} (\bibinfo {year}
  {1992})}\BibitemShut {NoStop}%
\bibitem [{\citenamefont {Morard}\ \emph {et~al.}(2020)\citenamefont {Morard},
  \citenamefont {Hernandez}, \citenamefont {Guarguaglini}, \citenamefont
  {Bolis}, \citenamefont {Benuzzi-Mounaix}, \citenamefont {Vinci},
  \citenamefont {Fiquet}, \citenamefont {Baron}, \citenamefont {Ko},
  \citenamefont {Gleason}, \citenamefont {Mao}, \citenamefont {Alonso-Mori},
  \citenamefont {Lee}, \citenamefont {Nagler}, \citenamefont {Galtier},
  \citenamefont {Sokaras}, \citenamefont {Glenzer}, \citenamefont {Andrault},
  \citenamefont {Garbarino}, \citenamefont {Mezouar}, \citenamefont
  {Schuster},\ and\ \citenamefont {Ravasio}}]{morard2020}%
  \BibitemOpen
  \bibfield  {author} {\bibinfo {author} {\bibfnamefont {G.}~\bibnamefont
  {Morard}}, \bibinfo {author} {\bibfnamefont {J.-A.}\ \bibnamefont
  {Hernandez}}, \bibinfo {author} {\bibfnamefont {M.}~\bibnamefont
  {Guarguaglini}}, \bibinfo {author} {\bibfnamefont {R.}~\bibnamefont {Bolis}},
  \bibinfo {author} {\bibfnamefont {A.}~\bibnamefont {Benuzzi-Mounaix}},
  \bibinfo {author} {\bibfnamefont {T.}~\bibnamefont {Vinci}}, \bibinfo
  {author} {\bibfnamefont {G.}~\bibnamefont {Fiquet}}, \bibinfo {author}
  {\bibfnamefont {M.~A.}\ \bibnamefont {Baron}}, \bibinfo {author}
  {\bibfnamefont {S.~H. S.~B.}\ \bibnamefont {Ko}}, \bibinfo {author}
  {\bibfnamefont {A.~E.}\ \bibnamefont {Gleason}}, \bibinfo {author}
  {\bibfnamefont {W.~L.}\ \bibnamefont {Mao}}, \bibinfo {author} {\bibfnamefont
  {R.}~\bibnamefont {Alonso-Mori}}, \bibinfo {author} {\bibfnamefont {H.~J.}\
  \bibnamefont {Lee}}, \bibinfo {author} {\bibfnamefont {B.}~\bibnamefont
  {Nagler}}, \bibinfo {author} {\bibfnamefont {E.}~\bibnamefont {Galtier}},
  \bibinfo {author} {\bibfnamefont {D.}~\bibnamefont {Sokaras}}, \bibinfo
  {author} {\bibfnamefont {S.~H.}\ \bibnamefont {Glenzer}}, \bibinfo {author}
  {\bibfnamefont {D.}~\bibnamefont {Andrault}}, \bibinfo {author}
  {\bibfnamefont {G.}~\bibnamefont {Garbarino}}, \bibinfo {author}
  {\bibfnamefont {M.}~\bibnamefont {Mezouar}}, \bibinfo {author} {\bibfnamefont
  {A.~K.}\ \bibnamefont {Schuster}},\ and\ \bibinfo {author} {\bibfnamefont
  {A.}~\bibnamefont {Ravasio}},\ }\bibfield  {title} {\bibinfo {title} {{In
  situ X-ray diffraction of silicate liquids and glasses under dynamic and
  static compression to megabar pressures}},\ }\href
  {https://doi.org/10.1073/pnas.192047011} {\bibfield  {journal} {\bibinfo
  {journal} {Proceedings of the National Academy of Sciences}\ }\textbf
  {\bibinfo {volume} {117}},\ \bibinfo {pages} {11981} (\bibinfo {year}
  {2020})}\BibitemShut {NoStop}%
\bibitem [{\citenamefont {Hemley}\ \emph {et~al.}(1988)\citenamefont {Hemley},
  \citenamefont {Jephcoat}, \citenamefont {Mao}, \citenamefont {Ming},\ and\
  \citenamefont {Manghnani}}]{hemley}%
  \BibitemOpen
  \bibfield  {author} {\bibinfo {author} {\bibfnamefont {R.~J.}\ \bibnamefont
  {Hemley}}, \bibinfo {author} {\bibfnamefont {A.~P.}\ \bibnamefont
  {Jephcoat}}, \bibinfo {author} {\bibfnamefont {H.~K.}\ \bibnamefont {Mao}},
  \bibinfo {author} {\bibfnamefont {L.~C.}\ \bibnamefont {Ming}},\ and\
  \bibinfo {author} {\bibfnamefont {M.~H.}\ \bibnamefont {Manghnani}},\
  }\bibfield  {title} {\bibinfo {title} {Pressure-induced amorphization of
  crystalline silica},\ }\href
  {https://doi.org/https://doi.org/10.1038/334052a0} {\bibfield  {journal}
  {\bibinfo  {journal} {Nature}\ }\textbf {\bibinfo {volume} {334}},\ \bibinfo
  {pages} {52} (\bibinfo {year} {1988})}\BibitemShut {NoStop}%
\bibitem [{\citenamefont {McQueen}\ and\ \citenamefont
  {Marsh}(1966)}]{Mcqueen}%
  \BibitemOpen
  \bibfield  {author} {\bibinfo {author} {\bibfnamefont {R.~G.}\ \bibnamefont
  {McQueen}}\ and\ \bibinfo {author} {\bibfnamefont {S.~P.}\ \bibnamefont
  {Marsh}},\ }\bibfield  {title} {\bibinfo {title} {Handbook of {Physical}
  {Constants} (unpublished data)},\ }\href@noop {} {\bibfield  {journal}
  {\bibinfo  {journal} {Geological Society of America Memoir}\ }\textbf
  {\bibinfo {volume} {97}},\ \bibinfo {pages} {153} (\bibinfo {year}
  {1966})}\BibitemShut {NoStop}%
\bibitem [{\citenamefont {Marsh}(1980)}]{marsh}%
  \BibitemOpen
  \bibfield  {author} {\bibinfo {author} {\bibfnamefont {S.}~\bibnamefont
  {Marsh}},\ }\bibfield  {title} {\bibinfo {title} {{LASL shock Hugoniot
  data}},\ }\href@noop {} {\bibfield  {journal} {\bibinfo  {journal}
  {University of California Press}\ }\textbf {\bibinfo {volume} {5}} (\bibinfo
  {year} {1980})}\BibitemShut {NoStop}%
\bibitem [{\citenamefont {Komabayashi}(2014)}]{komabayashi2014}%
  \BibitemOpen
  \bibfield  {author} {\bibinfo {author} {\bibfnamefont {T.}~\bibnamefont
  {Komabayashi}},\ }\bibfield  {title} {\bibinfo {title} {Thermodynamics of
  melting relations in the system {F}e-{F}e{O} at high pressure: Implications
  for oxygen in the {E}arth’s core},\ }\href
  {https://doi.org/https://doi.org/10.1002/2014JB010980} {\bibfield  {journal}
  {\bibinfo  {journal} {Journal of Geophysical Research: Solid Earth}\ }\textbf
  {\bibinfo {volume} {119}},\ \bibinfo {pages} {4164–4177} (\bibinfo {year}
  {2014})}\BibitemShut {NoStop}%
\end{thebibliography}%

\end{document}